\documentclass[showpacs,twocolumn,superscriptaddress]{revtex4}
\UseRawInputEncoding
\usepackage{doi}
\usepackage{hyperref}
\hypersetup{
  colorlinks=true,        
  linkcolor=blue,         
  citecolor=cyan,         
}

\usepackage[utf8x]{inputenc}
\DeclareUnicodeCharacter{2212}{\textendash}
\usepackage{graphicx}
\usepackage{dcolumn}
\usepackage{bm}
\usepackage{color}

\usepackage{amsmath}
\usepackage{amssymb}
\usepackage{mathrsfs}

\begin{document}

\title{Epicyclic motions and constraints on the charged stringy black hole spacetime }

\author{Sanjar Shaymatov}
\email{sanjar@astrin.uz}
\affiliation{Institute for Theoretical Physics and Cosmology, Zheijiang University of Technology, Hangzhou 310023, China}
\affiliation{Institute of Fundamental and Applied Research, National Research University TIIAME, Kori Niyoziy 39, Tashkent 100000, Uzbekistan}
\affiliation{National University of Uzbekistan, Tashkent 100174, Uzbekistan}
\affiliation{Power Engineering Faculty, Tashkent State Technical University, Tashkent 100095, Uzbekistan}

\author{Kimet Jusufi} \email{kimet.jusufi@unite.edu.mk
}
\affiliation{Physics Department, State University of Tetovo,
Ilinden Street nn, 1200, Tetovo, North Macedonia}

\author{Mirzabek Alloqulov}
\email{malloqulov@gmail.com}

\affiliation{Institute of Fundamental and Applied Research, National Research University TIIAME, Kori Niyoziy 39, Tashkent 100000, Uzbekistan}

\author{Bobomurat Ahmedov}
\email{ahmedov@astrin.uz}

 \affiliation{Institute of Fundamental and Applied Research, National Research University TIIAME, Kori Niyoziy 39, Tashkent 100000, Uzbekistan}
 \affiliation{Ulugh Beg Astronomical Institute, Astronomy St. 33, Tashkent 100052, Uzbekistan}
\affiliation{National University of Uzbekistan, Tashkent 100174, Uzbekistan}

\date{\today}
\begin{abstract}

The purpose of this paper is to examine the epicyclic motion of charged particles in the vicinity of a magnetically charged stringy black hole subject to an electromagnetic field. This investigation is motivated by the established fact that magnetic fields have a significant impact on the motion of charged particles in the vicinity of a black hole. It has been observed that the magnetic coupling parameter has a comparable impact on the motion of magnetic monopole charge, leading to an increase/decrease in the radius of the innermost stable circular orbit (ISCO) that plays a
crucial role in setting the accretion disk's the inner edge. Further, we analyze the characteristics of the epicyclic motion and determine  the generic form for the epicyclic frequencies. Our results demonstrate that the existence of the magnetic coupling parameter can account for the observed resonance of high frequency QPOs from selected three specific microquasars. We have applied the Monte Carlo simulations with the observed frequencies for three selected microquasars to constrain the parameters of the magnetically charged stringy black holes as well as coupling parameters that describe the interaction between the particle and the black hole. Finally, we study in detail the radiation properties of the accretion disk in the close vicinity of black hole. We show that the magnetic coupling parameter can have significant influence on the radiation flux of the accretion disk.

\end{abstract}
\pacs{
} \maketitle


\section{Introduction}
\label{introduction}

General Relativity (GR) has been regarded as the well accepted theory of gravity thus far in the strong field regime  regardless of the fact that there exist unresolved questions. In GR, one of the most intriguing events is that the gravitational collapse of massive stars at the end state of evolution is defined by fundamental mechanisms in which a black hole can be formed with its attractive and remarkable gravitational properties. However, black holes can be still predicted as a consequence of exact analytical solutions of the gravitational  field equations. 

Recent triumphal astronomical observations associated with the Event Horizon Telescope (EHT) and the LIGO-VIRGO collaborations have observed the shadow of supermassive black holes M87*~\cite{EHT19a,EHT19b}, SgrA* and gravitational waves (GW)~\cite{LIGO16a,LIGO16}, respectively, thus leading to the fact that for the first time GR has also been tested in the strong field regime. 

However, there are yet unsolved problems of GR, i.e., the inevitable occurrence of singularities, the spacetime quantization, etc. for which GR can lose its applicability. In this respect, the numerous alternative modified theories of gravity come into play in addressing these problems of GR as the potential candidates in the strong field regime. Therefore, it becomes particularly important to understand more deeply the nature and remarkable aspects of the existing gravitational fields, as well as their effects on the geodesics of test particles on the surrounding environment of the black holes. In this regard, recent modern observations would become a very important tool and platform to test not only the nature of the geometry but also the existing fields around black holes. Hence, these existing fields would become increasingly essential in affecting observable properties, i.e. the innermost stable circular orbit (ISCO) and astrophysical quasiperiodic oscillations (QPOs), etc.~\cite{Wald74,Benavides-Gallego19,Herrera00,Herrera05,Bini12,Toshmatov19d} and altering particle's geodesics in a variety of contexts~\cite{Shaymatov14,Shaymatov15,Dadhich18,Shaymatov19b,Shaymatov20egb,Shaymatov21c,Haroon19,Konoplya19plb,Hendi20,Jusufi19,Shaymatov21galaxy,Shaymatov21d,Rayimbaev-Shaymatov21a,Shaymatov21pdu}.

From an astrophysical point of view, the standard black holes can possess mass $M$, angular momentum $a$ and electric charge $Q$. If the black hole has a charge, and thus it refers  to Reissner-Nordstr\"{o}m (RN) black hole that can be characterized by mass and electric charge only. Regardless of the fact that black hole charge is still arguable RN black hole solution has so far been regarded as a very interesting solution~\cite{Zajacek19,Bally78,Pugliese11,Pugliese11b}. However, there exists an exact rotating Schwinger dyon solution that suggests that a black hole can possess  magnetic charge $Q_m$ as well~\cite{Kasuya82}. This solution is also a very fascinating solution with a magnetic charge. With this in view, many alternative solutions have been proposed so that black holes can be endowed with electric charge, as well as magnetic charge (see for example \cite{Kamata81,Shaymatov22a,Narzilloev20a,Narzilloev20b}). The point to be noted here is that the magnetic charge in nature is regarded as a hypothetical particle since it has not been observed yet. The existence of magnetic charge does however get predicted by string theory \cite{Wen85}. According to this theory, there exist distance interactions between particles so that they represent a particular behaviour of gravitational field assumed to be produced by the string elements accordingly~\cite{Letelier79}. Later on, following the Maxwell-Einstein theory magnetically charged RN black hole solution was proposed to be a new solution that generalizes electrically charged black hole solution in string theory~\cite{Gibbons88,Garfinkle91}. The most distinguishable aspect of such solutions is that magnetically charged black holes can be characterized by magnetic monopole charges \cite{Lee92PRL}. The magnetic monopole charge was initially proposed by Dirac that it would exist as a consequence of the electric charge quantization in the Universe~\cite{Dirac1931}. Taking it all together, the rotating charged black hole solution was obtained in heterotic string theory by Sen \cite{Sen92}, thereby referred to as the well known Kerr-Sen solution. A large amount of work has since been devoted to revealing the remarkable aspects of the Kerr-Sen black hole solution in a various contexts~\cite[see, e.g.,][]{Blaga01, An18,Gyulchev07,Younsi16,Hioki08,Dastan16,Uniyal18,Chen09b,
Larranaga11,Khani13,Siahaan15,Gonzalez17}. We noticed that there is also a recent investigation that shows the string effect on the relative time delay in the Kerr-Sen black hole solution~\cite{Izmailov20}.

It is well known that the astrophysical properties of gravitational compact objects can be analysed through the astronomical observations of particle outflows~\cite{Fender04mnrs,Auchettl17ApJ,IceCube17b} and the radiation emitted by an accretion disk. With this regard, analysing the rich observational phenomenology of electromagnetic radiation by a thin accretion disk with the expected thermal spectra can be used not only to provide information about gravity but also to test gravity in the strong field regime \cite{Abramowicz13}. For that, the electromagnetic field plays a crucial role in testing the above mentioned phenomena and the dynamics of charged
particles as well in the close vicinity of black holes. Therefore, in the present paper, we study the epicyclic motion with its applications to the quasiperiodic oscillations (QPOs) and the electromagnetic field effects on the particles moving in the disk around the magnetically charged stringy black hole with the aim of providing confidence in the conclusions of accretion disk observations. For that we consider QPOs observed in three GRO J1655-40, XTE J1550-564 and GRS 1915+105 microquasars in order to constrain black hole parameters and the charge coupling parameters and infer which values of these parameters best fit with the resonance of high frequency QPOs from three microquasars. We further investigate the properties of the accretion disk around the magnetically charged stringy black hole.    

As known black holes have been considered as a result of Einstein’s theory in GR and as a potential explanation for some X-ray and $\gamma$-ray observations. In this context, modern astronomical observations become increasingly important in testing the strong field regime of gravity and in constraining the precise parameters of black holes. With this in view, analysing QPOs with the X-ray data from the observations of accretion discs of astrophysical compact objects~\cite{Bambi12a,Bambi16b,Tripathi19} could play a decisive role in having information about unknown aspects of gravity and the geometry surrounding black holes. Note that QPOs can possess the X-ray power that has regularly been observed in microquasars being a primary source of low-mass binary systems that can have either neutron stars or black holes. QPOs have important astrophysical applications. Hence, the galactic microquasars  occur as a primary source of high-frequency (HF) QPOs that can be respectively defined by upper $\omega_{U}$ and lower $\omega_{L}$ frequencies with the ratio $\omega_{U}:\omega_{L}=3:2$ in most cases~\cite{Kluzniak01,Torok05A&A,Remillard06ApJ}. Particularly, such HF QPO models have so far been considered to address important aspects of the epicyclic motions on black hole accretion disks~\cite[see, e.g.,][]{Stuchlik13A&A,Stella99-qpo,Rezzolla_qpo_03a,Torok05A&A}. One can usually refer to such HF QPOs as the twin peak HF QPOs arising in pairs. This led to an increase in an activity where various models are proposed to address the QPOs in different settings \cite{Germana18qpo,Tarnopolski:2021ula,Dokuchaev:2015ghx,Kolos15qpo,Aliev12qpo,Stuchlik07qpo,Stuchlik21:Univ,Titarchuk05qpo,Rayimbaev-Shaymatov21a,Azreg-Ainou20qpo,Jusufi21qpo,Ghasemi-Nodehi20qpo,Rayimbaev22qpo} taking into account different types of resonances for the HF QPOs created in the accretion disk. There still yet remains open question about how HF QPOs occur in the accretion disk, i.e., no well-accepted model to explain such phenomenon (see for example~\cite{Torok11A&A}). For that there exist some alternative models proposed to settle this question with the help of an external magnetic field existing on the surrounding environment of black hole \cite{Tursunov20ApJ,Panis19,Shaymatov20egb,Shaymatov22c}.  

We organize this paper as follows: In
Sec.~\ref{Sec:metic} we discuss a magnetically charged black hole spacetime metric with its properties and then consider the charged particle motion orbiting around black hole. In Sec.~\ref{Sec:qpo} we explore and compare epicyclic motions with the observed resonances  of HF QPOs for three selected microquasars in the surrounding environment of black hole, which is followed by the main discussion of constraints on the parameters of the magnetically charged stringy black holes, as well as the charge coupling parameters in Sec.~\ref{Sec:Constrain}.  In Sec.~\ref{Sec:accretion} we study the accretion disk around the magnetically charged stringy black hole. We end up with conclusion in Sec.~\ref{Sec:Conclusion}.


\section{Magnetically charged stringy black hole and charged test particles dynamics \label{Sec:metic} }

We explore spacetime of a magnetically charged stringy black hole for which the action in the heterotic string theory is written by \cite{Horowitz93,Gonzalez17}
\begin{eqnarray}\label{Eq:action}
\mathcal{S}&=& \int d^4 x \sqrt{-g}\ e^{-2\varphi} \Big[ R+4(\nabla \varphi)^2 -
F_{\alpha\beta}F^{\alpha\beta}  \nonumber\\ &-&\frac{1}{12} H_{\alpha\beta\gamma}
H^{\alpha\beta\gamma}\Big]\, .  
\end{eqnarray}
In the above action, $\varphi$ comes into play as the dilaton field, while $e^\varphi$ represents a coupling constant. Note that $H_{\alpha\beta\gamma}$ is written in terms of potential  $B_{\alpha\beta}$ and the Maxwell gauge field $A_\alpha$ as 
\begin{eqnarray}
H_{\alpha\beta\gamma}&=&\partial_{\alpha} B_{\beta\gamma}+\partial_{\gamma} B_{\alpha\beta}+\partial_{\beta} B_{\gamma\alpha}\nonumber\\& -& \frac{1}{4}\Big(A_{\alpha} F_{\beta\gamma}+A_{\gamma} F_{\alpha\beta}+A_{\beta} F_{\gamma\alpha}\Big)\, .
\end{eqnarray}
For the standard Einstein-Hilbert action one has to set $H_{\alpha\beta\gamma}=0$.  It turns out that the action Eq.~(\ref{Eq:action}) can be simplified by rescaling the metric tensor $g_{\alpha\beta}\rightarrow e^{-2\varphi} g_{\alpha\beta}$, that is given by  
\begin{eqnarray}\label{Eq:action1}
S=\int d^4x\ \sqrt{-g}\ \left(R - 2(\nabla\varphi)^2 - e^{-2\varphi}
F^2\right)\, , 
\end{eqnarray}
for which the equations of motion for Maxwell field can be written by the rescaling factor 
\begin{eqnarray}\label{Eq:eq-motion}
\nabla_\alpha\left(e^{-2\varphi} F^{\alpha\beta}\right) = 0\, . \end{eqnarray}
The above equation remains invariant for the following transformations $F \rightarrow F^{\star}$ and
$\varphi\rightarrow -\varphi$ and satisfies $F^{\star}_{\alpha\beta} = e^{-2\varphi} \frac{1}{2}
{\epsilon_{\alpha\beta}}^{\gamma\rho}F_{\gamma\rho}$, thus resulting in allowing electrically charged black hole solution to be transformed into a magnetically charged black hole solution under the electromagnetic duality transformation $\varphi\rightarrow -\varphi$~(see for example ~\cite{Garfinkle91,Gonzalez17}). In doing so, a magnetically charged black hole solution is given by the following line element:  
\begin{eqnarray}\label{metric2}
 ds^2= -\frac{f(r)}{h(r)}dt^2+\frac{dr^2}{f(r)h(r)}+r^2d\theta^2+ r^2\sin^2\theta\,d\phi^2\, ,
\end{eqnarray}
where $f(r)$ and $h(r)$ are defined as 
\begin{eqnarray}
f(r)=1-\frac{2\, M}{r}\, , \mbox{~~}h(r)=1-\frac{Q^2_m}{M\, r}\, ,
\end{eqnarray}
where $M$ and $Q_m$,  respectively, refer to black hole mass and its magnetic charge. 

Hereafter we consider both electrically and magnetically charged particles dynamics around magnetically charged stringy black hole. We assume that charged test particle has the rest mass $m$ and electric and magnetic charges $q$ and $q_m$, respectively. For that we first write the Hamiltonian of the system as~\cite{Misner73}  
\begin{eqnarray}
 H  \equiv  \frac{1}{2}& g^{\alpha\beta}&\left(\frac{\partial \mathcal{S}}{\partial
x^{\alpha}}-q{A}_{\alpha}+iq_m{A}^{\star}_{\alpha}\right)\nonumber\\  &\times &\left(\frac{\partial \mathcal{S}}{\partial
x^{\beta}}-q{A}_{\beta}+iq_m{A}^{\star}_{\beta}\right)\, ,
\label{Eq:H}
\end{eqnarray}
where $S$ and $x^{\alpha}$ respectively refer to the action and the four-vector coordinate, while $A_{\alpha}$ and $A^{\star}_{\alpha}$ correspond to the vector and the dual vector potentials of the electromagnetic field. The non-vanishing components of these vector potentials are given by 
\begin{eqnarray}\label{4pots}
A_{\alpha}&=&\left(-0,0,0,-Q_m\cos\theta\right)\, , \\  
A^{\star}_{\alpha}&=&\left(-\frac{iQ_m}{r},0,0,0\right)\, .
\end{eqnarray}
\begin{figure*}
\begin{center}
\begin{tabular}{c c}
  \includegraphics[scale=0.6]{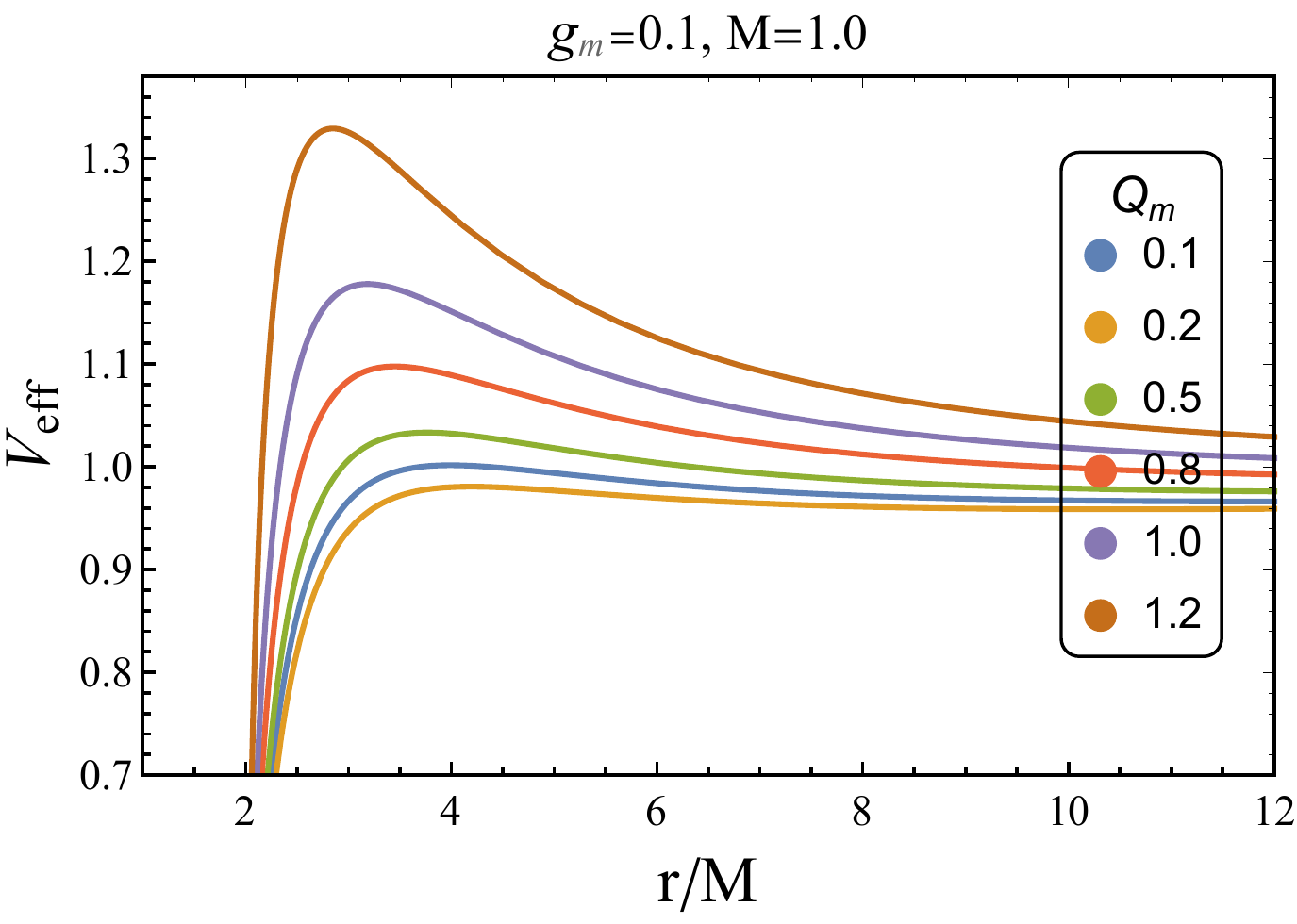}\hspace{0cm}
  \includegraphics[scale=0.6]{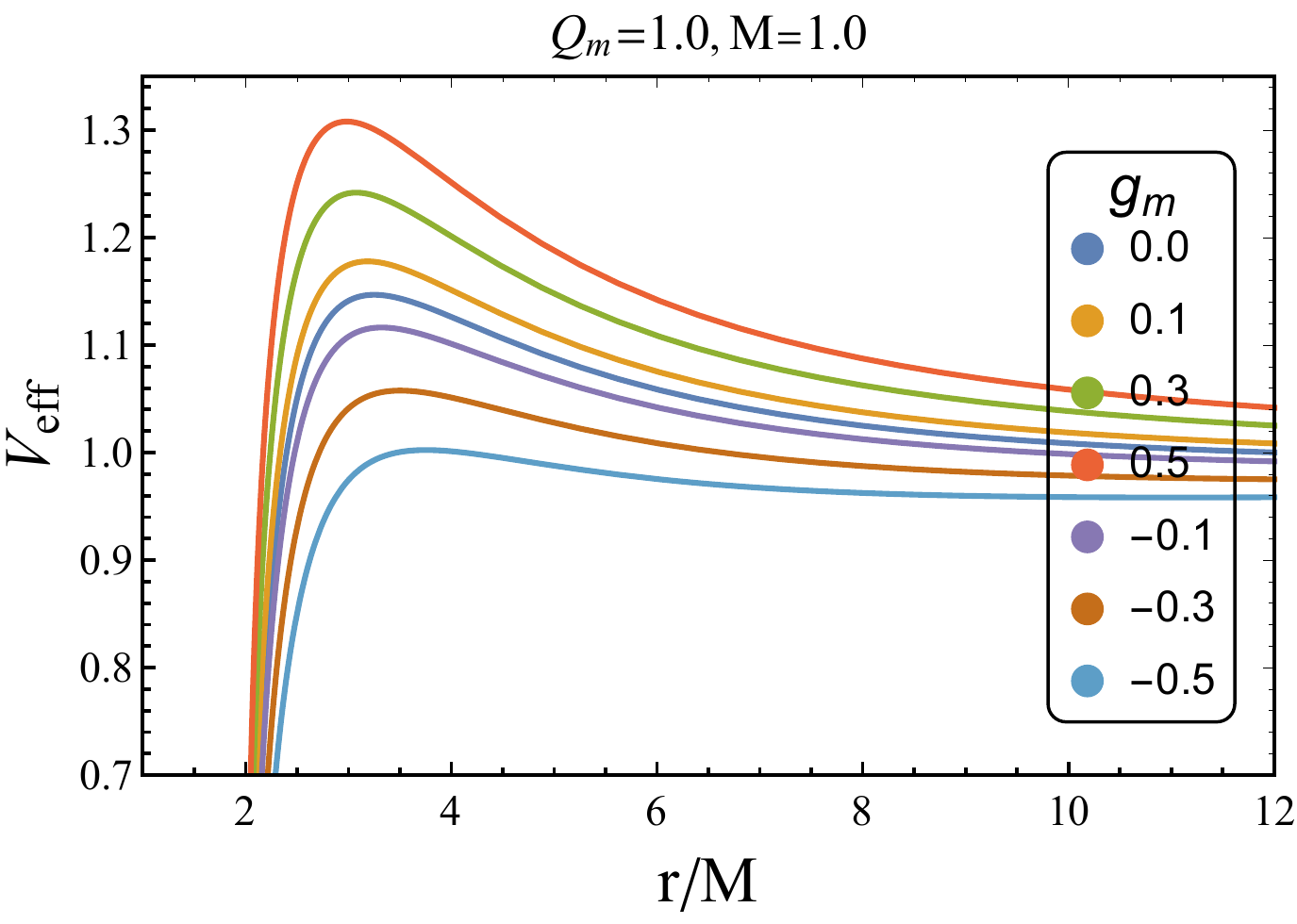}\hspace{0cm}
\end{tabular}
	\caption{\label{fig:eff_pot} 
Radial profile of the effective potential for the radial motion of test particles orbiting around the magnetically charged stringy black hole.  Left panel: $V_{eff}$ is plotted for various values of black hole magnetic charge $Q_m$ while keeping fixed $g_m$. Right panel: $V_{eff}$ is plotted for various values of the charge coupling parameter $g_m$ while keeping fixed $Q_m$.}
\end{center}
\end{figure*}

Let us then consider the Hamiltonian of the system, i.e., $H=k/2$ along with $k=-m^2$. In the Hamilton-Jacobi equation, the action $S$ is written by
\begin{eqnarray}\label{Eq:sep1}
\mathcal{S}= -\frac{1}{2}k\lambda-Et+L\varphi+\mathcal{S}_{r}(r)+\mathcal{S}_{\theta}(\theta)\ ,
\end{eqnarray}
with $S_{r}$ and $S_{\theta}$ that we define later as the functions of $r$ and $\theta$, respectively. Following to Eq.~(\ref{Eq:sep1}) the Hamilton-Jacobi equation can be rewritten as follows:  
\begin{eqnarray}\label{Eq:sep}
&-&
\frac{h(r)}{f(r)}\left[-E+
\frac{q_m\,Q_m}{r} \right]^{2}
+f(r)~h(r)
\left(\frac{\partial \mathcal{S}_{r}}{\partial
r}\right)^{2}\nonumber\\&+&{1\over r^{2}}\left(\frac{\partial \mathcal{S}_{\theta}}{\partial
\theta}\right)^{2}+\frac{\left(L+qQ_m\cos\theta\right)^{2}}{r^{2} \sin^2\theta }-k=0 \, . 
\end{eqnarray}
As can be seen from the separated form of the Hamilton-Jacobi equation there appear four conserved quantities of motion; specific energy $E$, angular momentum $L$, and $k$  ~\cite{Misner73}. However, the fourth one is associated with the latitudinal motion that can be omitted for the particle motion that takes place at the equatorial (i.e., $\theta=\pi/2$). For further analysis we shall for simplicity separate Eq.~(\ref{Eq:sep}) into two dynamical $H_{\textrm{dyn}}$ and potential $H_{\textrm{pot}}$ parts, which read as follows: 
\begin{eqnarray}
H_{\textrm{dyn}}&=& \frac{1}{2}\left[\frac{h(r)}{f(r)^{-1}} \left(\frac{\partial S_{r}}{\partial r}\right)^2 +\frac{1}{r^2}\left(\frac{\partial S_{\theta}}{\partial \theta}\right)^2\right],\\
H_{\textrm{pot}}&=& \frac{1}{2}\left[\frac{\left(-{E}+\frac{q_mQ_m}{r}\right)^2}{f(r)\,h(r)^{-1}}+\frac{\left({L}+qQ_{m}\cos\theta\right)^2}{r^2\sin^2\theta}
-{k}\right]\, . \nonumber\\\label{Eq:sepham}
\end{eqnarray}
\begin{figure*}
\begin{tabular}{cccc}
  \includegraphics[scale=0.45]{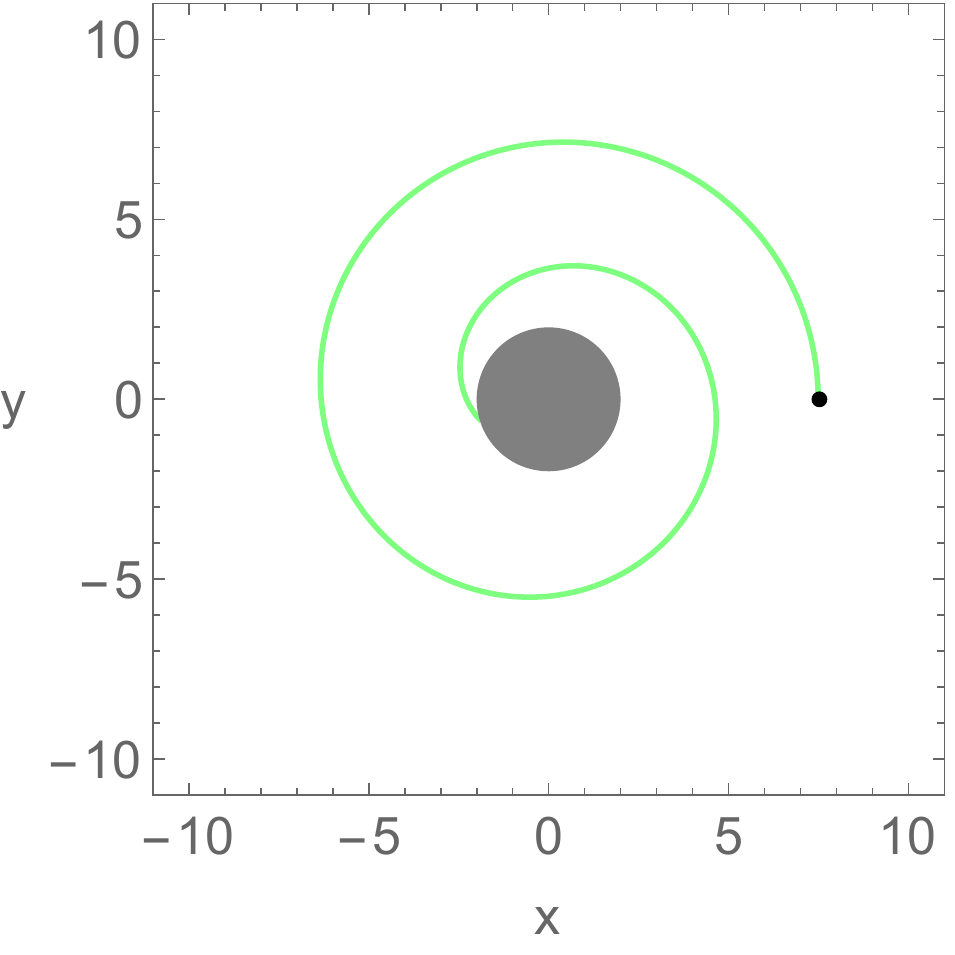}\hspace{-0.0cm}
  \includegraphics[scale=0.45]{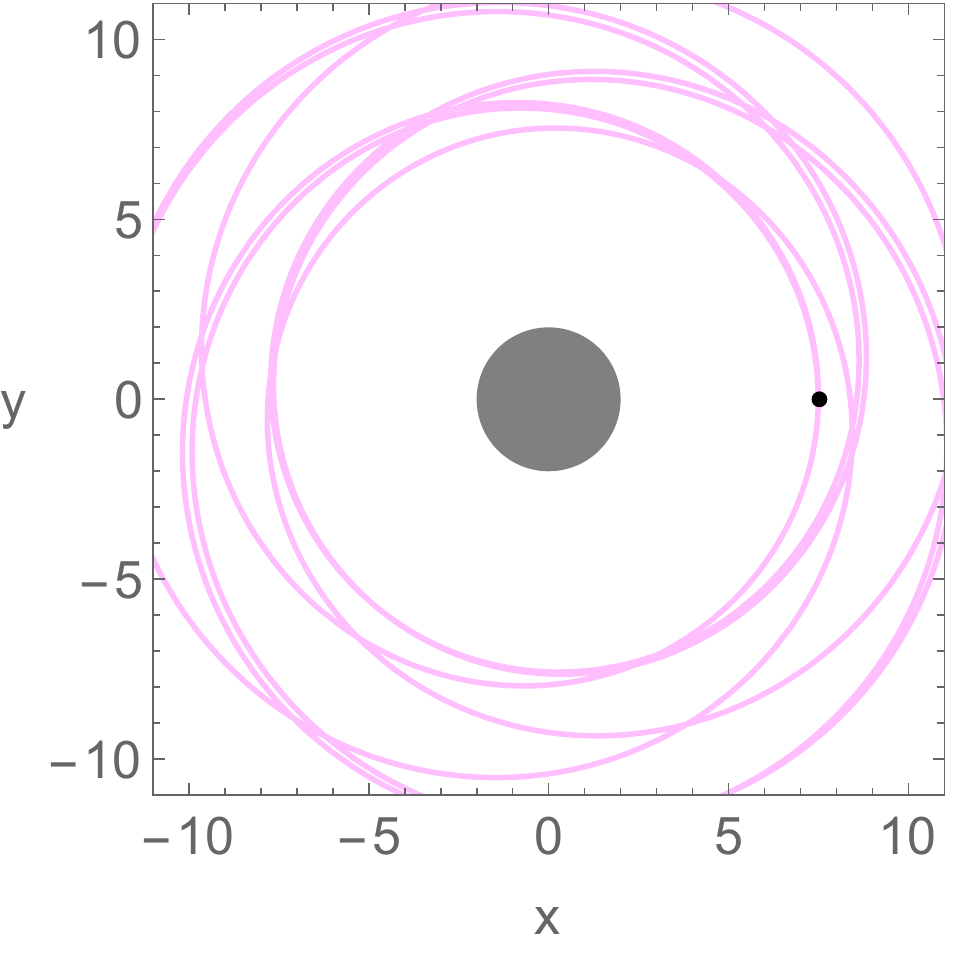}
  \includegraphics[scale=0.45]{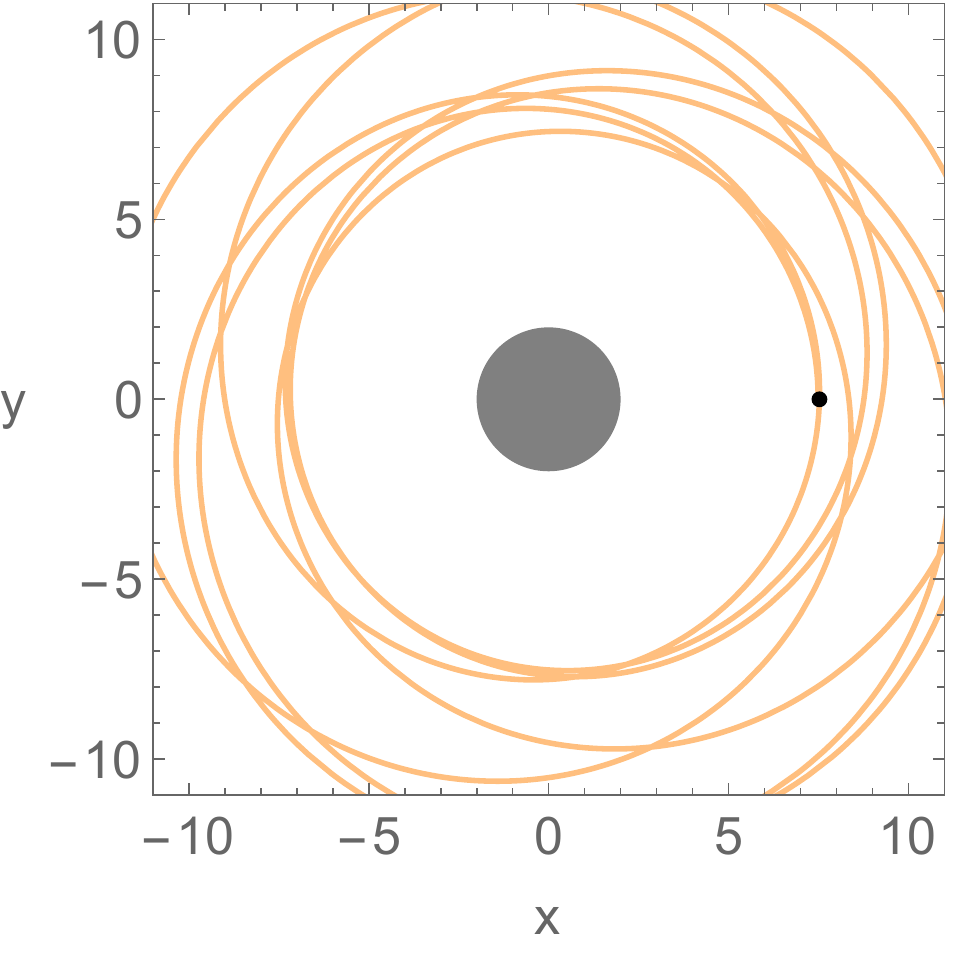}
  \includegraphics[scale=0.45]{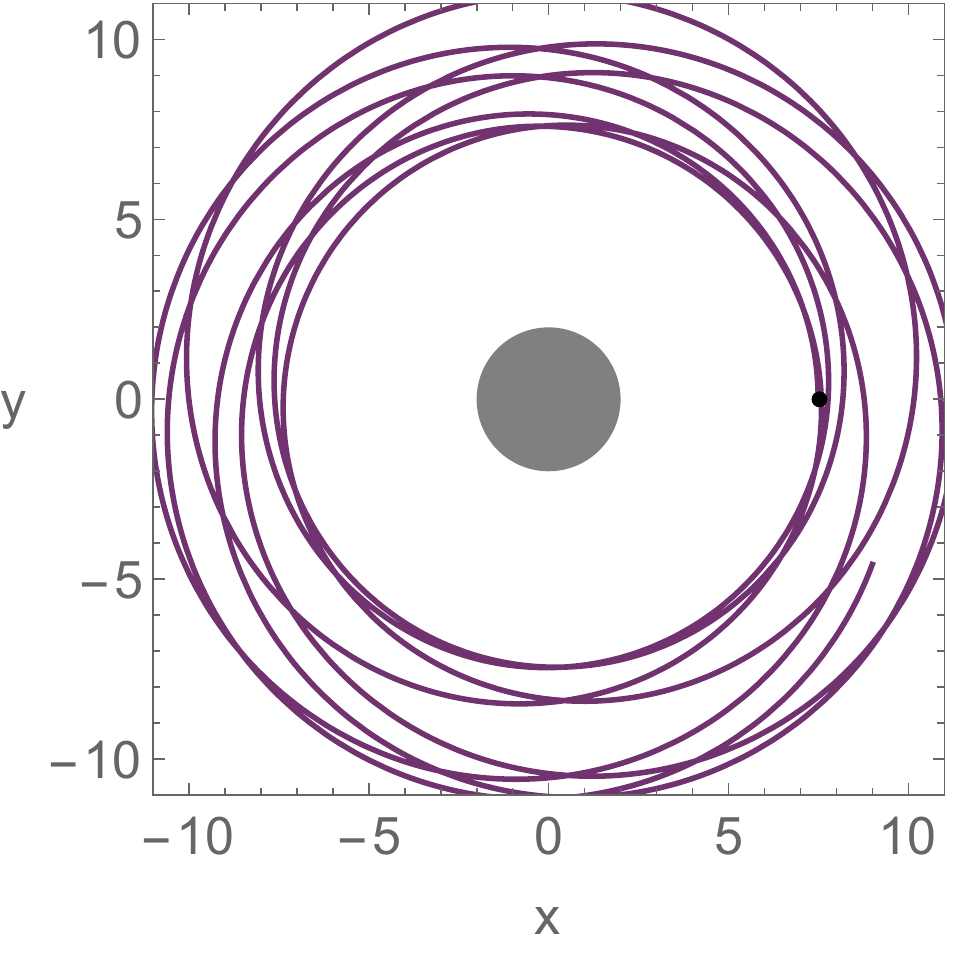}\\
  
  \includegraphics[scale=0.45]{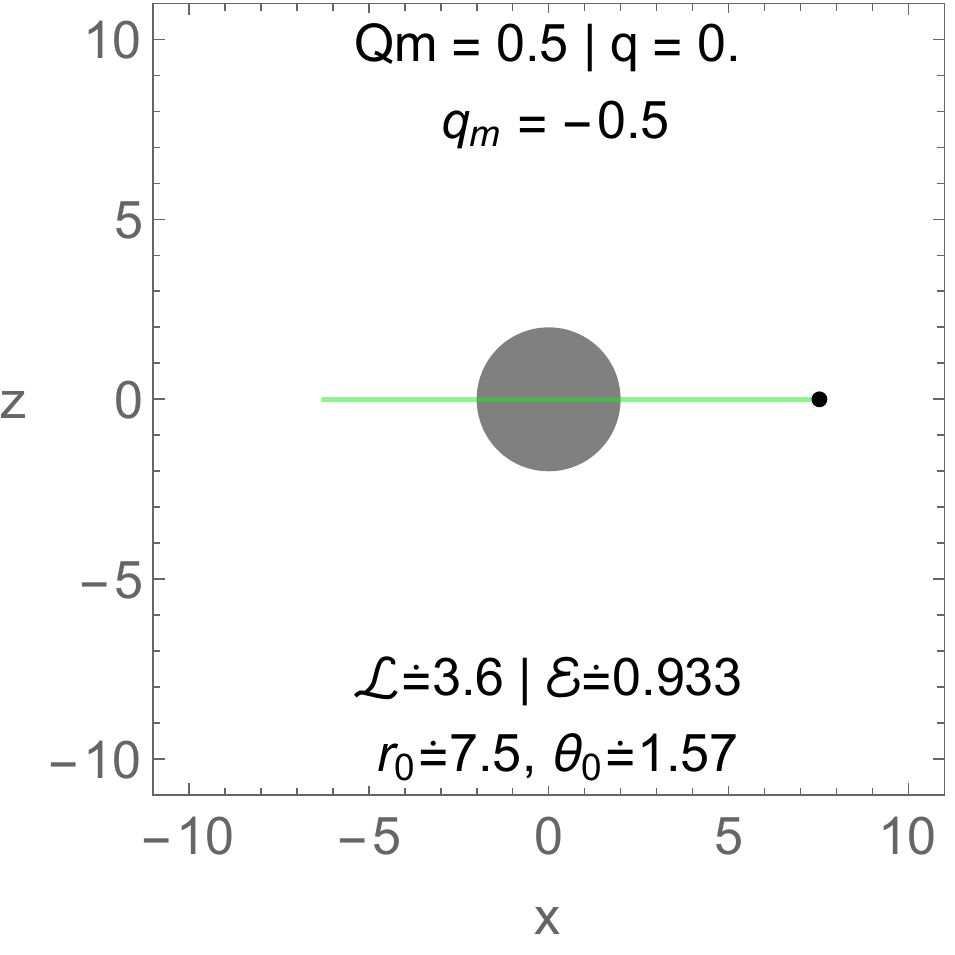}\hspace{0.0cm}
  \includegraphics[scale=0.45]{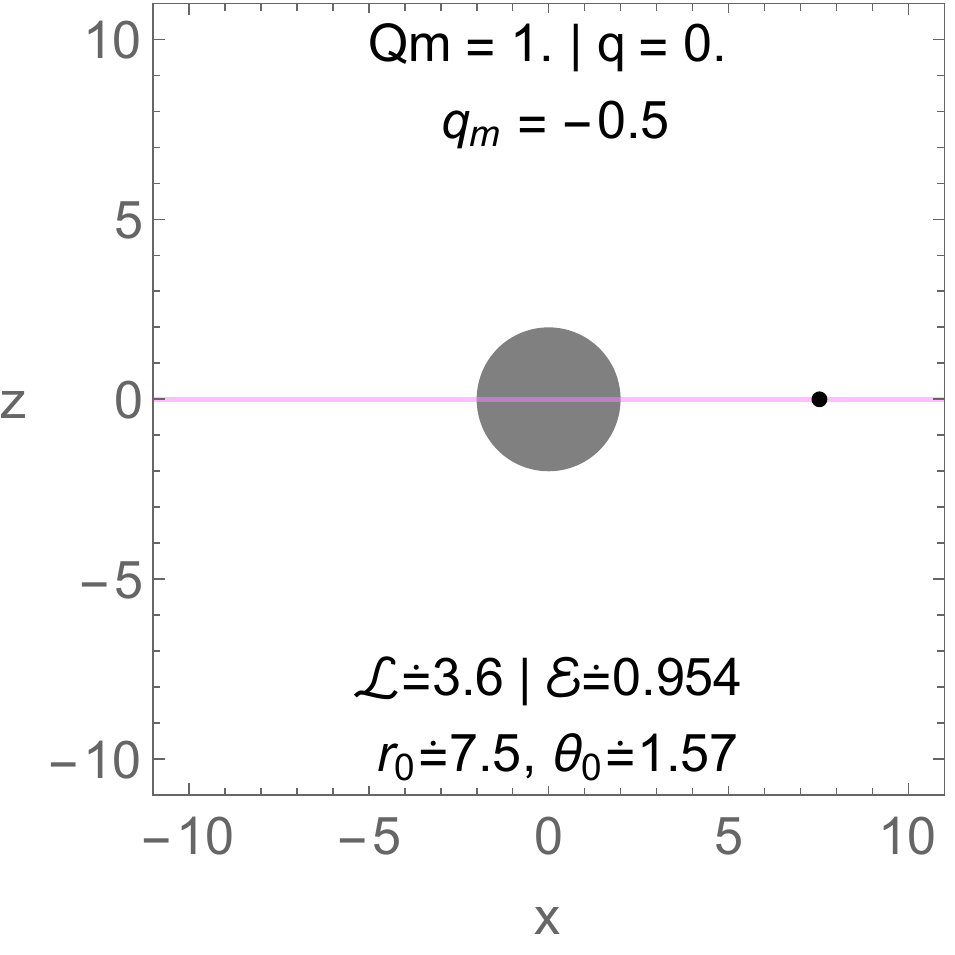}
  \includegraphics[scale=0.45]{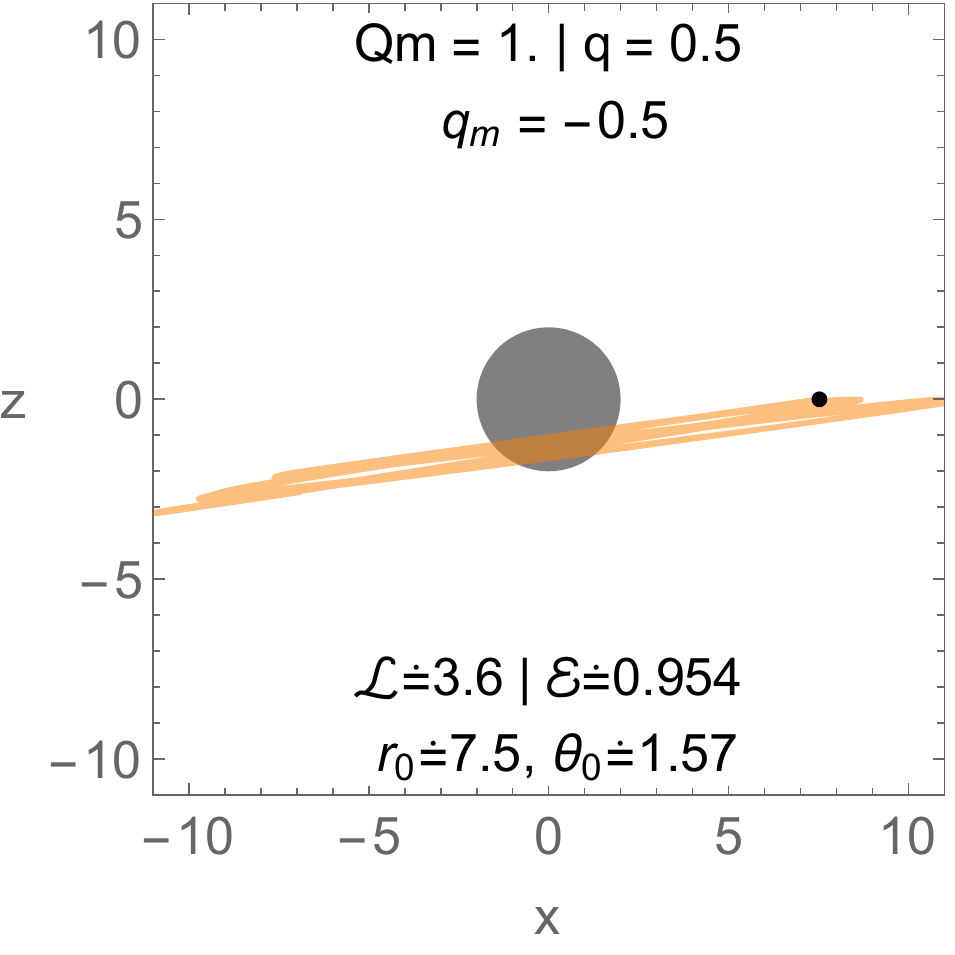}
  \includegraphics[scale=0.45]{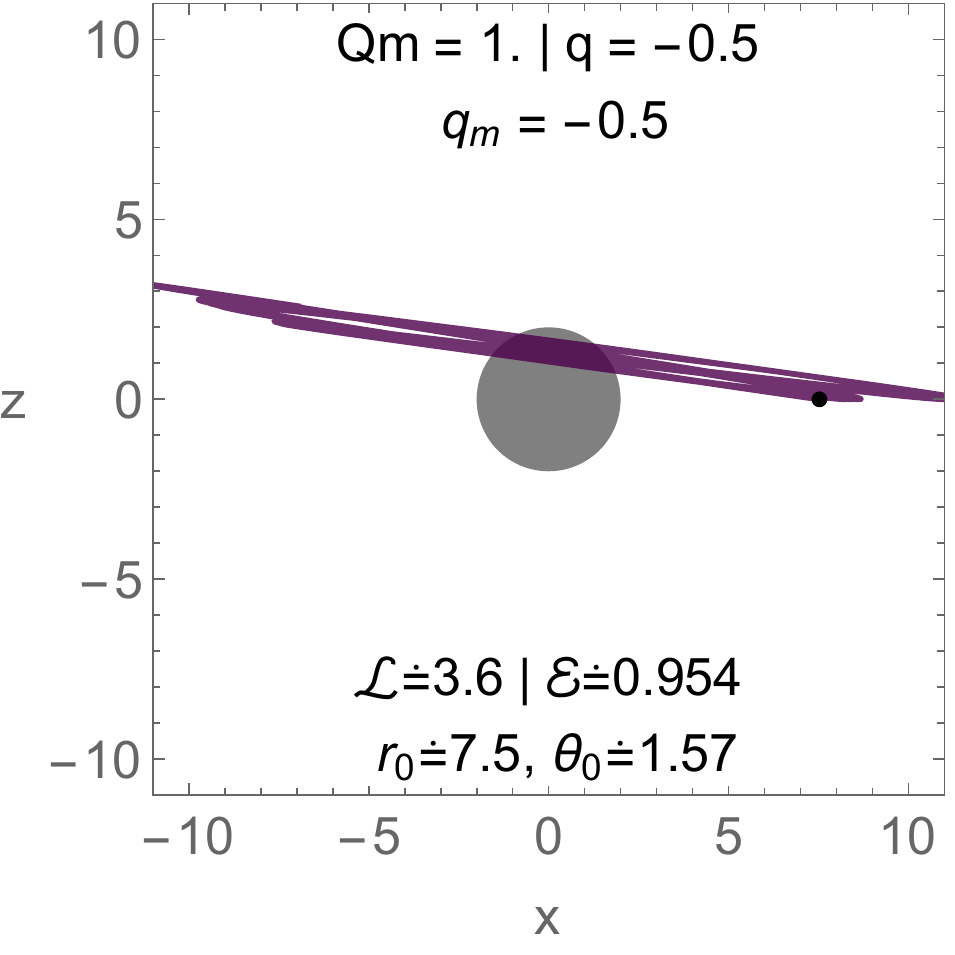}
  \end{tabular}
	\caption{\label{fig:traj2D} 
2D trajectories of magnetically and electrically charged particles orbiting the magnetically charged stringy black hole. Top panel reflects the behaviour of trajectories that are observed from the polar view, i.e. $z = 0$, while the bottom panel reflects the behaviour of trajectories from the equatorial view, i.e. $y = 0$, for various possible cases. }
\end{figure*}
\begin{table*}
\begin{center}
\caption{Numerical values of the  
{ISCO parameters $\mathcal{L}_{ISCO}, \mathcal{E}_{ISCO}, r_{ISCO}$, $v_{ISCO}$ and $\Omega_{ISCO}$} of the test particles orbiting on the ISCO radius around the magnetically charged stringy black hole are tabulated for various possible cases. Note that we set $\theta=\pi/2$}\label{tab:table1}
\resizebox{\linewidth}{!}
{
\begin{tabular}{l l |l l l l l| l l l l l}
 \hline \hline
 & & &\multicolumn{2}{c}{} & $g_{m}=0.2$ & &\multicolumn{2}{c}{} & $g_{m}=0.4$ & &  
 \\
\cline{3-7}\cline{8-12}
& $Q_{m}$    & $\mathcal{L}_{ISCO}$ & $\mathcal{E}_{ISCO}$ & $r_{ISCO}$ & $v_{ISCO}$ & $\Omega_{ISCO}$ & $\mathcal{L}_{ISCO}$ & $\mathcal{E}_{ISCO}$ & $r_{ISCO}$ & $v_{ISCO}$ & $\Omega_{ISCO}$  \\
\hline
& 0.1   & 3.12120 & 0.95446  & 6.02699 & 0.49748 & 0.06752 & 2.76342 & 0.96635  & 6.13410 & 0.49099 & 0.06576  \\
& 0.5   & 2.92772 & 0.96138  & 6.09328 & 0.47213 & 0.06485 & 2.56444 & 0.97371  & 6.32282 & 0.45907 & 0.06125 \\
& 0.8     & 2.56597 & 0.97288 & 6.26331 & 0.42149 & 0.05859 & 2.17622 & 0.98566 & 6.90408 & 0.39093 & 0.05010 \\
& 1.0       & 2.15476 & 0.98370 & 6.58936 & 0.35838 & 0.04928 & 1.67931 & 0.999600 & 8.82341 & 0.28747 & 0.03042 \\
\hline
& &&\multicolumn{2}{c}{}&  $g_{m}=-0.2$ && \multicolumn{2}{c}{} & $g_{m}=-0.4$ & &  
\\
\cline{3-7}\cline{8-12}
& $Q_{m}$    & $\mathcal{L}_{ISCO}$ & $\mathcal{E}_{ISCO}$ & $r_{ISCO}$ & $v_{ISCO}$ & $\Omega_{ISCO}$ & $\mathcal{L}_{ISCO}$ & $\mathcal{E}_{ISCO}$ & $r_{ISCO}$ & $v_{ISCO}$ & $\Omega_{ISCO}$\\  
\hline
& 0.1   & 3.77589 & 0.93219 & 6.01522 & 0.49821 & 0.06772 & 4.08373 & 0.92183 & 6.05513 &0.49575 & 0.06705 \\
& 0.5   & 3.58031 & 0.93793 & 5.97260 & 0.47945 & 0.06688 & 3.88453 & 0.92693 & 5.98198 &0.47887 & 0.06672 \\
& 0.8     & 3.22560 & 0.94745 & 5.87819 & 0.44357 & 0.06493 & 3.52566 & 0.93530 & 5.82584 & 0.44684 & 0.06588 \\
& 1.0       & 2.84346 & 0.95644 & 5.74187 & 0.40224 & 0.06223 & 3.14256 & 0.94299 & 5.61353 & 0.41031 & 0.06468 \\
 \hline \hline
\end{tabular}
}
\end{center}
\end{table*}

As was mentioned earlier the radial $S_{r}$ and angular $S_{\theta}$ parts of the Hamiltonian can be separable and thus they are given by
\begin{align}
&S_r=\int\sqrt{\left(E-\frac{q_mQ_m}{r}\right)^2-\frac{f(r)}{h(r)}\left(-k+\frac{K}{r^2}\right)}\,\frac{dr}{f(r)}\ , \label{HJ1} \\ &S_\theta=\int \sqrt{K-\left(\frac{L+qQ_m\cos\theta}{\sin\theta}\right)^2}\,d\theta\ , \label{HJ2}
\end{align}
with $K$ referred to as the well-known Carter constant. For further analysis we shall for convenience define 
\begin{eqnarray}
{\cal E}&=&\frac{E}{m}\ ,~ {\cal L}=\frac{L}{mM}\ ,~ {\cal K}=\frac{K}{(mM)^2}\ , ~\frac{k}{m^2}=-1\, ,
\end{eqnarray}
and as well as we normalize the radial coordinate as $r\to r/M$.  Following to the Hamiltonian of the system $m dx^\alpha/d\lambda=g^{\alpha\beta}\left(\partial S/\partial x^\beta-qA_{\beta}+iq_mA_{\beta}^{\star}\right)$ along with an affine parameter, $\lambda$, the equations of motion can be written as follows: 
\begin{eqnarray}\label{Eq:tdot}
\dot{t}&=&\frac{h(r)}{f(r)}\left({\cal E}-\frac{g_m}{r}\right)\ ,\\\dot{\phi}&=&\frac{{\cal L}+\sigma_m \cos\theta}{r^2\sin^2\theta}\ ,\\
\label{Eq:rdot}\dot{r}^2&=&h(r)^2\Bigg[\left({\cal E}-\frac{g_m}{r}\right)^2-\frac{f(r)}{h(r)}\left(1+\frac{\cal K}{r^2}\right)\Bigg]
\geq 0\ , \\
\label{Eq:thetadot}\dot{\theta}^2&=&\frac{1}{r^4}\left[{\cal K}-\left(\frac{{\cal L}+\sigma_m\cos\theta}{\sin\theta}\right)^2\right]
\geq 0\ ,
\end{eqnarray}
where we introduce the following notations 
\begin{align}
\sigma_m=\frac{qQ_m}{mM}\ \qquad\mbox{and} \qquad g_m=\frac{q_mQ_m}{mM}\, , 
\end{align}
and we further refer them as the charge coupling parameters, i.e, referred to as electric and magnetic coupling parameters, respectively. From the procedure above the timelike radial motion of charged particles orbiting magnetically charged stringy black hole is generically given by 
\begin{eqnarray}\label{Eq:rdot}
\dot{r}^2=\Big[\mathcal{E}-\mathcal{E}_-(r)\Big]\Big[\mathcal{E}-\mathcal{E}_+(r)\Big]\, ,
\end{eqnarray}
where $\mathcal{E}_{\pm}(r)$ for the motion in the equatorial plane is defined by 
\begin{eqnarray} \label{Eq:Veff2}
\mathcal{E}_{\pm}(r)=\frac{g_{m}}{r} \pm
\sqrt{\frac{f(r)}{h(r)}\left(1+\frac{\mathcal{L}^2}{r^{2}}\right)}\, .
\end{eqnarray}
From Eqs.~(\ref{Eq:rdot}) and (\ref{Eq:Veff2}),   the following condition, either $\mathcal{E}>\mathcal{E}_{+}(r)$ or $\mathcal{E}<\mathcal{E}_{-}(r)$, must be satisfied so that $\dot{r}^2\geq 0$ always. However, we focus on $\mathcal{E}_{+}(r)$ since it is more relevant to the effective potential. Therefore, we shall further restrict ourselves to the case in which $\mathcal{E}_{+}(r)=V_{\rm eff}(r)$ that we show in Fig.~\ref{fig:eff_pot}. In Fig.~\ref{fig:eff_pot}, we demonstrate the radial profile of the effective potential, $V_{\rm eff}$, for the radial motion of magnetically charged particle. As can be seen from Fig.~\ref{fig:eff_pot}, the left panel shows the effect of black hole's magnetic charge $Q_m$ on the behaviour of the effective potential for the fixed coupling parameter $g_m$, while the right panel represents the impact of the coupling parameter $g_m$ for its various combinations in the case of fixed $Q_m$. From the left panel, as one increases black hole magnetic charge the curves of the effective potential shift upward, thereby resulting in reinforcing gravitational barrier. Similarly to what is observed for black hole magnetic charge, the curves the effective potential get shifted upward as a consequence of an increase in the positive value of $g_m$. Meanwhile they go down as $g_m<0$ increases, in turn  weakening gravitational potential; see Fig.~\ref{fig:eff_pot}. It is worth noting that circular orbits existing around the black hole shift towards left to smaller $r$ for $g_m<0$, while towards right to larger $r$ for $g_m>0$, resulting in pushing it away from the black hole horizon.  Two dimensional trajectories of charged particles are also reflected in Fig.~\ref{fig:traj2D}, leading to  qualitative analysis of the behaviour of charged particles orbiting around the black hole as a consequence of the existing forces. In \ref{fig:traj2D}, we demonstrate 2D trajectories, observed from the polar view (i.e. $z = 0$) and the equatorial view (i.e. $y = 0$), for magnetically and electrically charged particles moving at the orbits restricted to the equatorial plane of the magnetically charged stringy black hole. As can be seen from \ref{fig:traj2D}, we focus on the captured and bound orbits so as to reveal the impact of black hole magnetic charge and test particle charges. From the top row of Fig.~\ref{fig:traj2D}, it is shown that orbits become bound orbits, initially captured ones, as $Q_m$ increases in the case of fixed magnetic charge $q_m$, whereas from the bottom row the inclusion of electric charge $q$ makes the bound orbits occur either above or below the equatorial plane, depending on its negative and positive charge. One can infer from the particle trajectories that they become increasingly important to explain the behaviour of bound and captured orbits around black hole under the combined effect of parameter $Q_m$, electric $q$ and magnetic charge $q_m$ of the particle.
\begin{figure*}
\begin{center}
\begin{tabular}{c c}
  \includegraphics[scale=0.6]{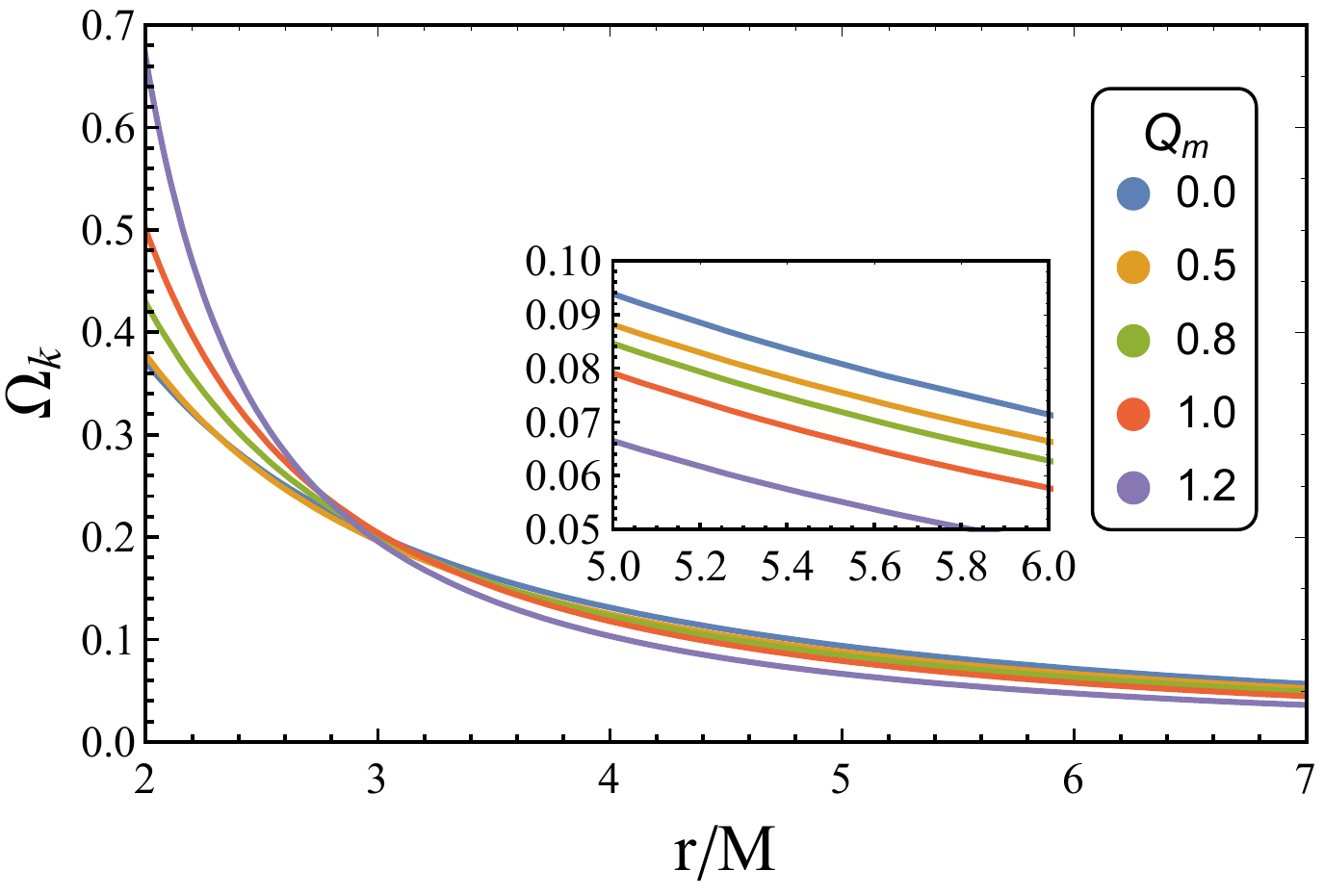}\hspace{0cm}
 & \includegraphics[scale=0.6]{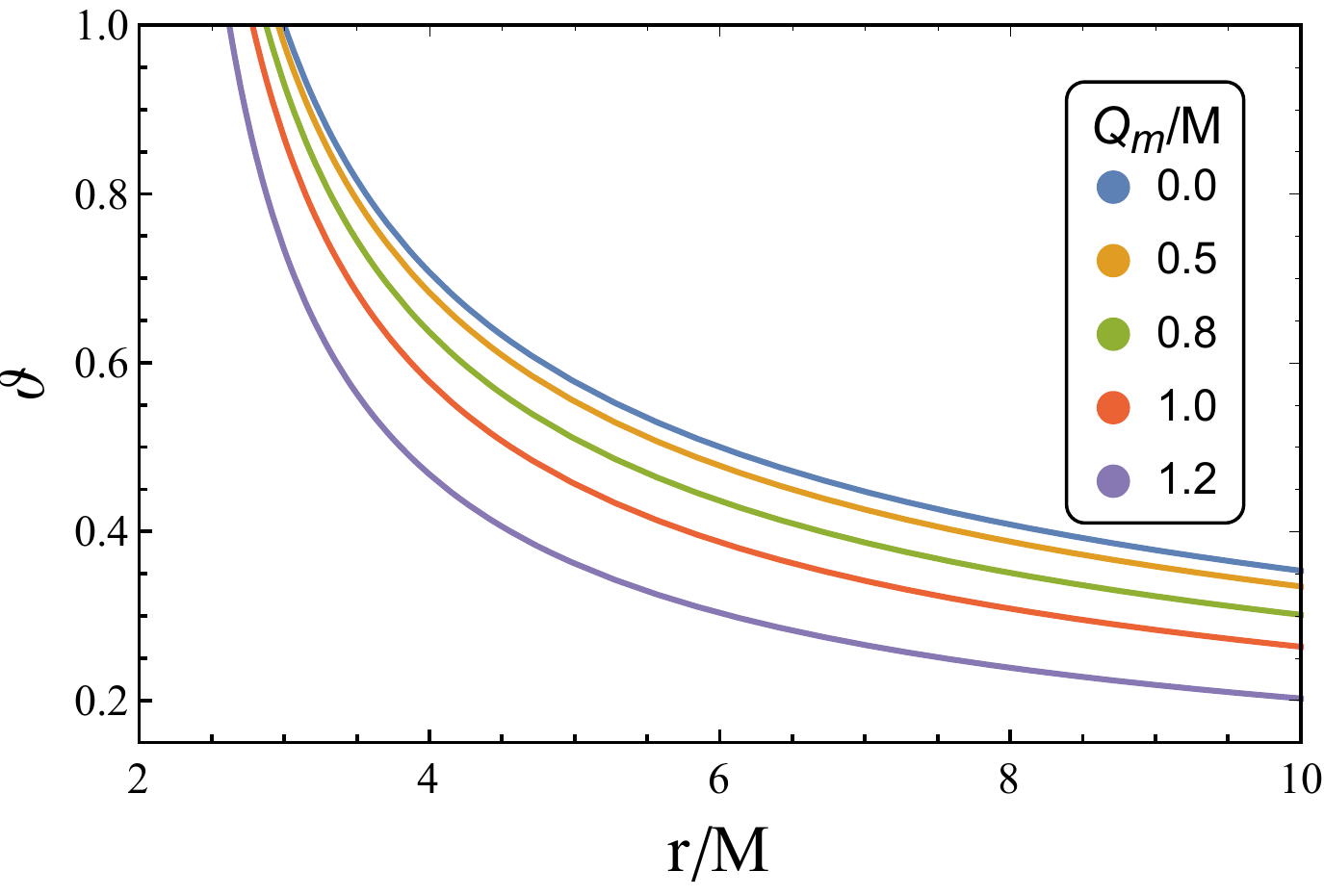}
\end{tabular}
	\caption{\label{fig:freq_vel} 
Radial profile of the orbital velocity $v$ is plotted for particles orbiting charged stringy black hole for various combinations of magnetic charge $Q_m$. }
\end{center}
\end{figure*}

Next, let's focus our attention on the circular orbits surrounding a black hole. The condition for charged particles to be at the circular orbits is given by the effective potential and its first derivative  
\begin{eqnarray}\label{Eq:circular}
V_{\rm eff}(r)=\mathcal{E} \mbox{~~and~~}  V_{\rm eff}(r)_{,r}=0\, ,
\end{eqnarray}
where $_{,r}$ denotes a derivative with respect to $r$. Taking into account above condition one can obtain specific energy $\mathcal{E}$ and angular momentum $\mathcal{L}$ for charged particles to move in circular orbits. Specific energy $\mathcal{E}$ is determined by Eq.~(\ref{Eq:Veff2}), while angular momentum $\mathcal{L}$ reads as
\begin{eqnarray}
\mathcal{L}^2&=& \frac{2 g_m^2 F+2 r^3 F F_{,r}- r^4F_{,r}^2}{r^2 4F^2-4 r F F_{,r}+F_{,r}^2}\nonumber\\&+&\frac{2 \sqrt{-2 g_m^2 r^3 F^2 F_{,r}+g_m^4 F^2+4 g_m^2 r^2 F^3}}{4 F^2-4 r F F_{,r}+r^2 F_{,r}^2}\, ,
\end{eqnarray}
where we have defined $F=f(r)/h(r)$. Note that for this angular momentum $\mathcal{L}$ there is still an extremum in the effective potential for particle to move in circular orbits. It is however increasingly important to determine the innermost stable circular orbit (ISCO) for which the maximum and minimum of effective potential coincide. Therefore, it is required to have one more condition for determining the ISCO radius of the particle, that is given by the second derivative of $V_{eff}$ with respect to $r$ 
    \begin{eqnarray}       V_{eff(r)_{,rr}}=0\, .\label{Eq:ddVeff}
    \end{eqnarray}
We now explore the ISCO parameters, $\mathcal{L}_{ISCO}, \mathcal{E}_{ISCO}$ and $r_{ISCO}$, numerically and tabulate their numerical values in Table \ref{tab:table1} for various cases. As can be seen from \ref{tab:table1} the ISCO radius increases as a consequence of an increase in the value of $Q_m$. Hence, one can infer that black hole magnetic charge, $Q_m$, comes into play as an attractive gravitational charge. Note that, similarly to what is observed for $Q_m$ the charge coupling parameter has a similar effect when considering its positive values $g_m>0$, whereas it has an opposite effect when considering its negative values $g_m<0$, i.e., the ISCO radius gets decreased, as seen in Table \ref{tab:table1}. However, the combined effect of both $g_{m}$ and $Q_{m}$ generally increases the ISCO radius. Furthermore, $\mathcal{L}_{ISCO}$ gets increased and $\mathcal{E}_{ISCO}$ gets decreased, respectively, as a consequence of the combined effect of both $g_{m}>0$ and $Q_{m}$. Meanwhile, the opposite is true for the combined effect of $g_{m}<0$ and $Q_{m}$. Also one can see that as $Q_m$ we increases $\mathcal{E}_{ISCO}$ increases and $\mathcal{L}_{ISCO}$ decreases accordingly so that the particle moves in the ISCO.         
 
Now we consider the radial and angular velocity of the particle orbiting on the ISCO around magnetically charged black hole, similarly to what was observed for $\mathcal{E}_{ISCO}$ and  $\mathcal{L}_{ISCO}$. For that we recall Eqs.~(\ref{Eq:tdot})-(\ref{Eq:thetadot}) so that we introduce the coordinate velocity that is measured by a local observer~\cite{Shapiro83,Misner73}. The coordinate velocity components are then written as follows: 
\begin{align}\label{vr}
&v_{\hat r}=\sqrt{-\frac{g_{rr}}{g_{tt}}}\frac{dr}{dt}=\sqrt{1-\frac{f(r)}{h(r)}\frac{r^2+{\cal K}}{(r{\cal E}-g_m)^2}}\ , \\
&v_{\hat\theta}=\sqrt{-\frac{g_{\theta\theta}}{g_{tt}}}\frac{d\theta}{dt}=\frac{\sqrt{f(r)/h(r)}}{r{\cal E}-g_m}\sqrt{{\cal K}-\left(\frac{{\cal L}+\sigma_m\cos\theta}{\sin\theta}\right)^2}\ ,\\
&v_{\hat\phi}=\sqrt{-\frac{g_{\phi\phi}}{g_{tt}}}\frac{d\phi}{dt}=\frac{\sqrt{f(r)/h(r)}}{r{\cal E}-g_m}\frac{{\cal L}+\sigma_m \cos\theta}{\sin\theta}\, .\label{vf}
\end{align}
The point to be noted here is that we can omit the parameter ${\cal K}$ provided that the motion is restricted to the equatorial plane (i.e., $\theta=\pi/2$). From Eqs.~(\ref{vr})-(\ref{vf}) it is straightforward to obtain the classical form of the particle energy as 
\begin{eqnarray}
{\cal E}=\frac{\sqrt{f(r)/h(r)}}{\sqrt{1-v^2}}+\frac{g_m}{r}\  \mbox{\, with \, } v^2=v_{\hat r}^2+v_{\hat\theta}^2+v_{\hat\phi}^2\ .    
\end{eqnarray}
It is vital to note that near the horizon, $f(r)=0$, the radial velocity can be defined by $v_{\hat r}=1$, whereas for the rest components $v_{\hat\theta}=v_{\hat\phi}=0$ is always satisfied. As was mentioned above it would become increasingly important to determine linear velocity for particles moving in the ISCO. From this perspective we define orbital velocity $v_\phi$, that is given by (\cite{Pugliese11,Shaymatov22a}) 
\begin{align}
v=v_\phi=\Omega\sqrt{-\frac{g_{\phi\phi}}{g_{tt}}}\ ,  \end{align}
with $\Omega$ being the orbital angular velocity for the particle, i.e. the Keplerian angular frequency. Note that the rest velocities vanish at the ISCO, i.e. $v_r=v_{\theta}=0$. 
However, one can generally obtain $\Omega$ in the charged particle case, thus following the non-geodesic equation that can be written as follows: 
\begin{eqnarray}
g_{tt,r}+\Omega^2g_{\phi\phi,r}&=&-\frac{2q}{m}\frac{\Omega A_{\phi,r}}{\sqrt{-g_{tt}-\Omega^2g_{\phi\phi}}}\, .
\end{eqnarray}
The above equation can be solved to give 
\begin{eqnarray}\label{Eq:kep}
\Omega^2&=&\Bigg[{\Omega_{0}^2}-2{g_{tt}}\left(\frac{qA_{\phi,r}}{mg_{\phi\phi,r}}\right)^2\pm 
\frac{2qA_{\phi,r}}{mg_{\phi\phi,r}} \nonumber\\&\times&\left\{-g_{tt}\Omega_{0}^2 -g_{\phi\phi}\Omega_{0}^4+\left(\frac{qA_{\phi,r}\,g_{tt}}{mg_{\phi\phi,r}}\right)^2 \right\}^{1/2}\Bigg]\nonumber\\&\times& \left[1+4g_{\phi\phi}\left(\frac{qA_{\phi,r}}{mg_{\phi\phi,r}}\right)^2\right]^{-1}\, .
\end{eqnarray}
The above equation can be reduced to  $\Omega^2=\Omega_{0}^2={-g_{tt,r}/g_{\phi\phi,r}}$ for the particle, $q=0$. The same is also case for the Keplerian frequency in the case in which $A_{\phi}=-Q_m\cos\theta$. Therefore, Eq.~(\ref{Eq:kep}) yields  
\begin{eqnarray}
\Omega_{k}=\sqrt{\frac{M \left(2 M^2-Q_m^2\right)}{2 r \left(Q_m^2-M r\right)^2\sin^2\theta}}\, .
\end{eqnarray}
Taking altogether the orbital velocity at the equatorial plane can be written as  
{\begin{align}\label{eq:v_r}
v=\sqrt{\frac{1}{2}\Big[\frac{2M}{r-2 M}+\frac{Q^2_{m}}{ Q^2_{m}- M r}\Big]}\, .
\end{align}}
Hereafter we explore $\Omega_k$ and $v$ at the ISCO radius numerically. Note that Eq.~(\ref{eq:v_r}) recovers $v_{ISCO}$ for the Schwarzschild black hole case in the case in which $Q_{m}=0$, i.e., $v_{ISCO}=1/2$. 

In Fig.~\ref{fig:freq_vel} we show the radial profile of the Keplerian frequency $\Omega_{k}$ and orbital velocity $v$ of particles moving in the circular orbits around magnetically charged stringy black hole, measured by a local observer. As can be easily seen from Fig.~\ref{fig:freq_vel} the Keplerian frequency decreases at larger $r$ as we increase $Q_m$, while the opposite is the case at smaller $r$, especially very close to the black hole
horizon.  The left panel of Fig.~\ref{fig:freq_vel} reflects an impact of $Q_m$ on the radial profile of orbital velocity.  This clearly shows that orbital velocity for test particle moving in the circular orbits gets decreased as $Q_m$ increases, thereby resulting in shifting curves down to smaller $v$. One can however state that both $\Omega_k$ and $v$ mainly increase for test particles when approaching to smaller $r$. To be more quantitative, we analysed both $\Omega_k$ and $v$ by exploring numerically once more, thus leading to understand their behaviour more deeply at the ISCO radius. The numerical values of both $\Omega_{ISCO}$ and $v_{ISCO}$ are also tabulated in Table \ref{tab:table1}. As seen in Table \ref{tab:table1}, one can deduce that the Keplerian frequency $\Omega$ and the orbital velocity $v$ respectively decrease as $Q_m$ increases similarly to what is observed for the angular velocity $\mathcal{L}$ for the particle moving in the ISCO around the charged stringy black hole. Also we notice that the negative coupling charge parameter $g_m<0$ has an opposite effect on $\Omega$ and $v$ in contrast to its positive one; see Table \ref{tab:table1}.

\section{\label{Sec:qpo}
The frequencies of epicyclic motion}

\begin{figure}
\begin{tabular}{c }
  \includegraphics[scale=0.6]{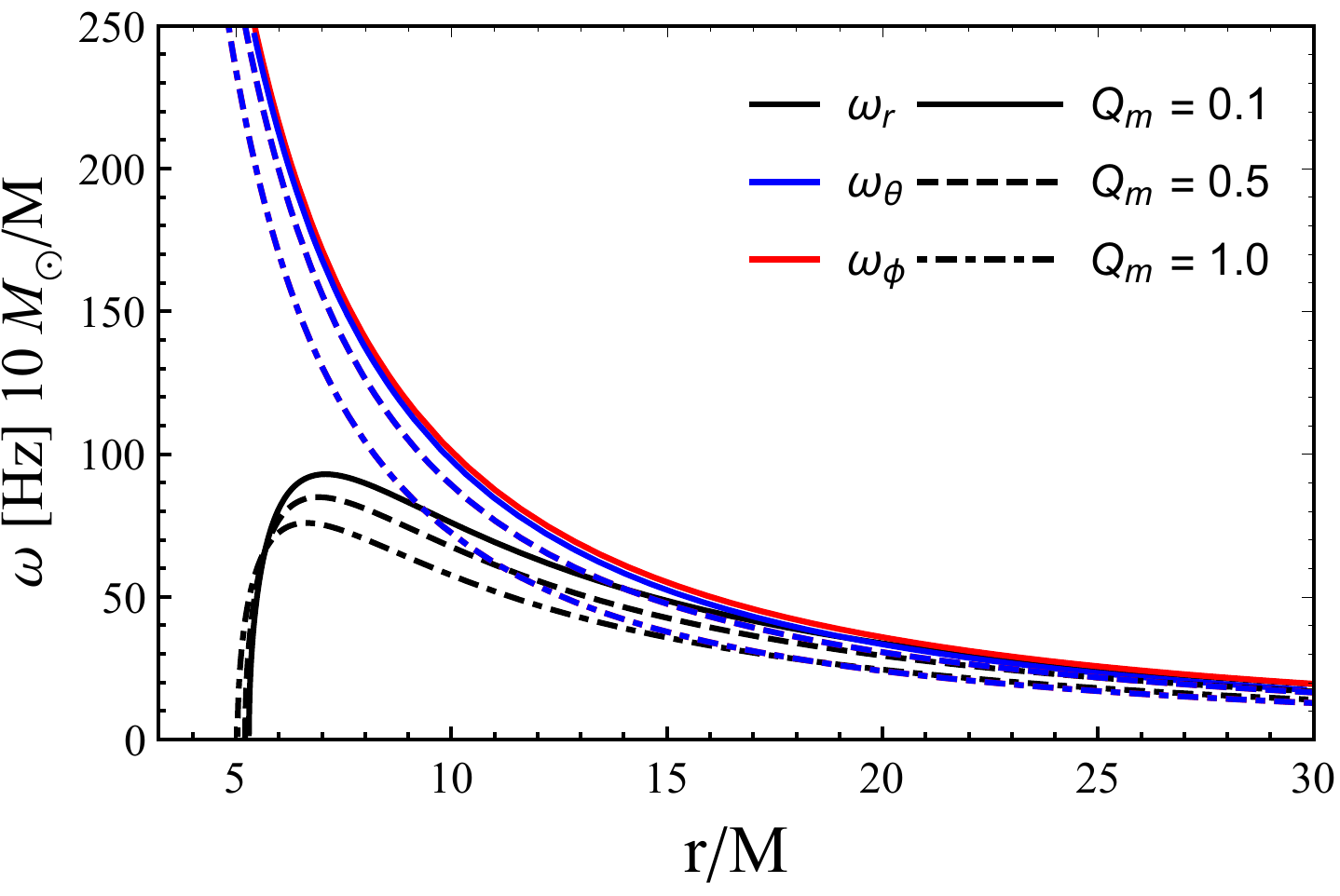}\hspace{-0.0cm}
  
\end{tabular}
\caption{\label{fig:qpo_freq} The epicyclic frequencies $\omega_r$, $\omega_{\theta}$ and $\omega_{\phi}$ as a function of $r/M$ are plotted for a neutral particle case for various combinations of black hole magnetic charge $Q_m$. }
\end{figure}
\begin{figure*}
\begin{tabular}{c c c}
  \includegraphics[scale=0.4]{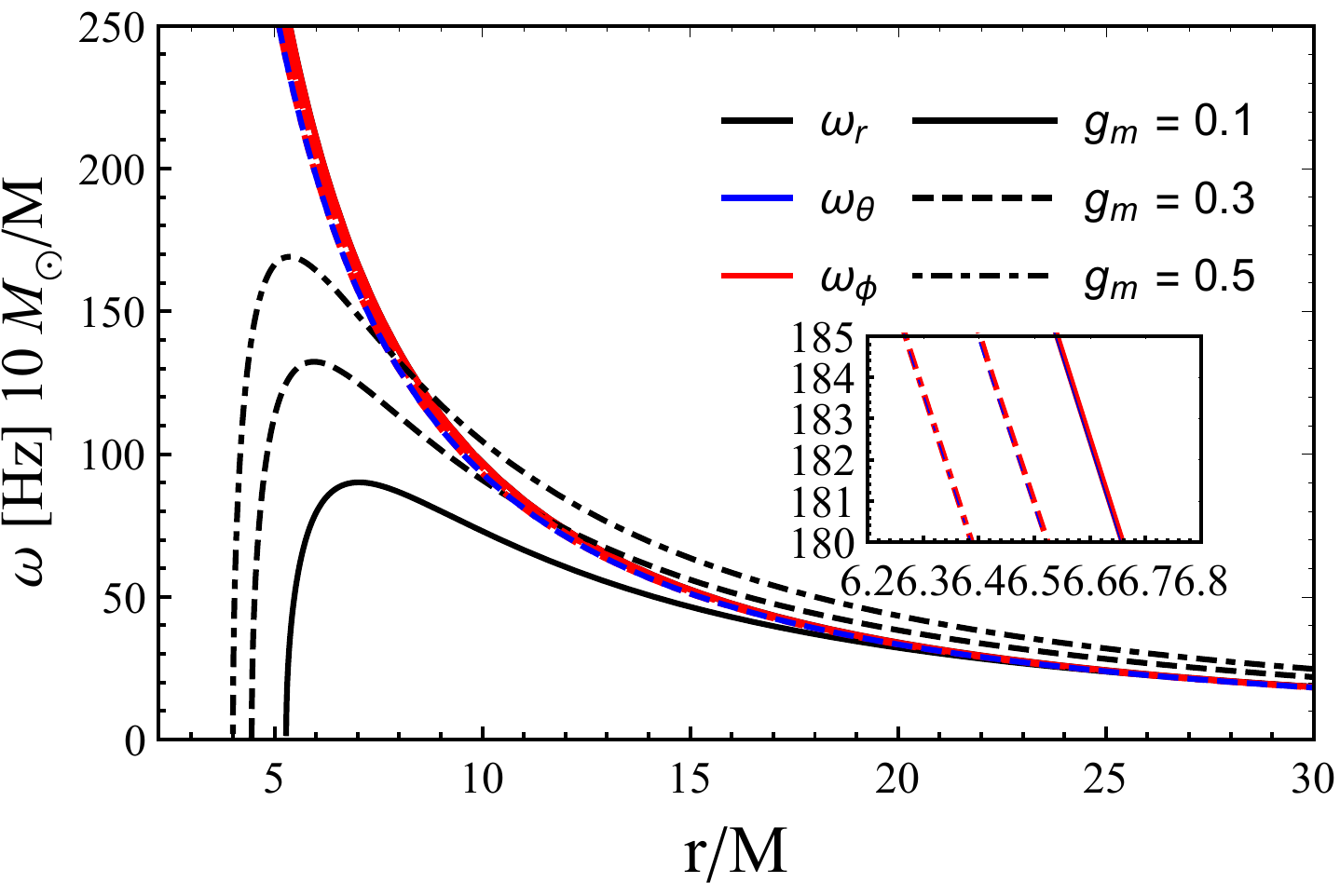}\hspace{-0.0cm}
   &  \includegraphics[scale=0.4]{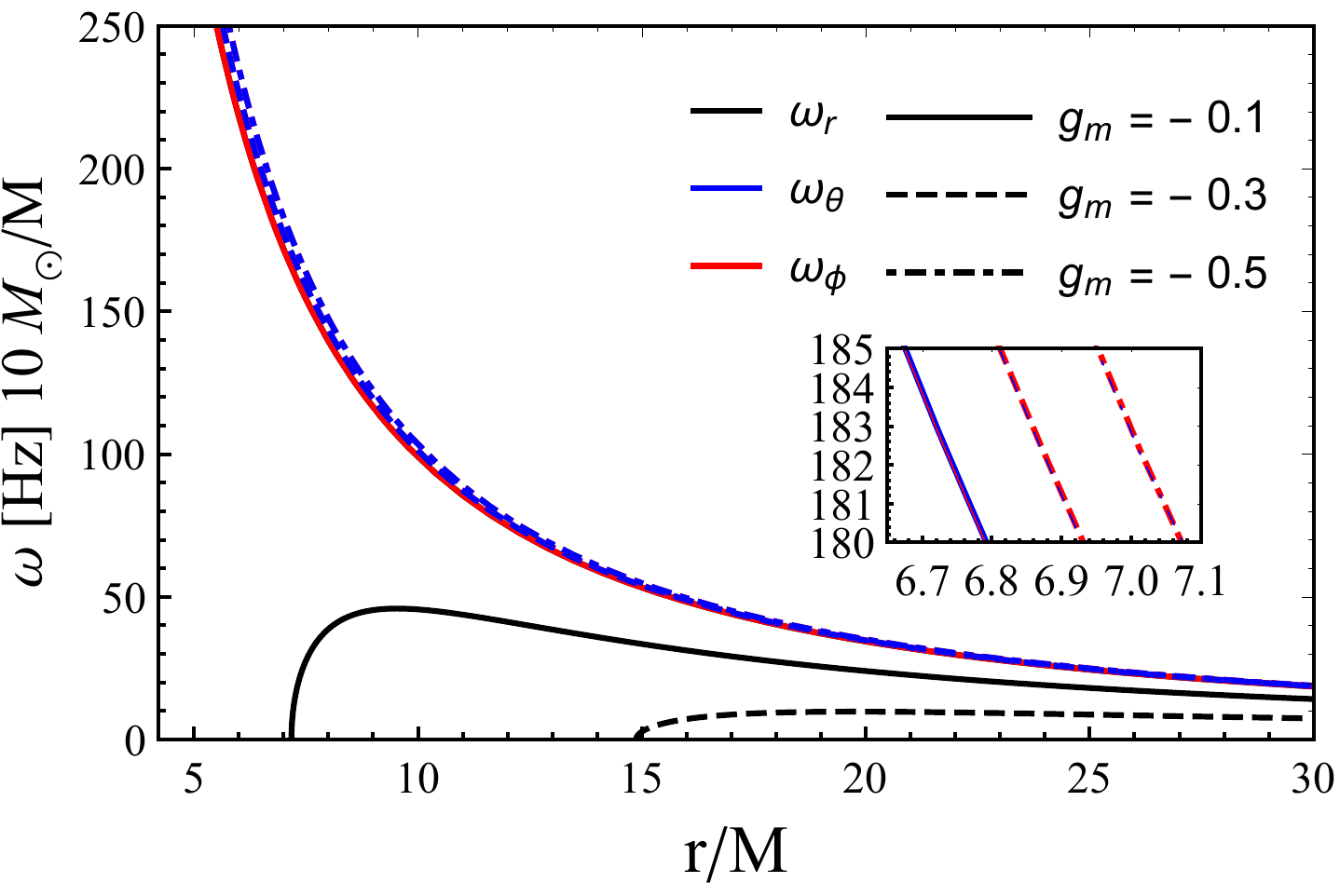}
   \includegraphics[scale=0.4]{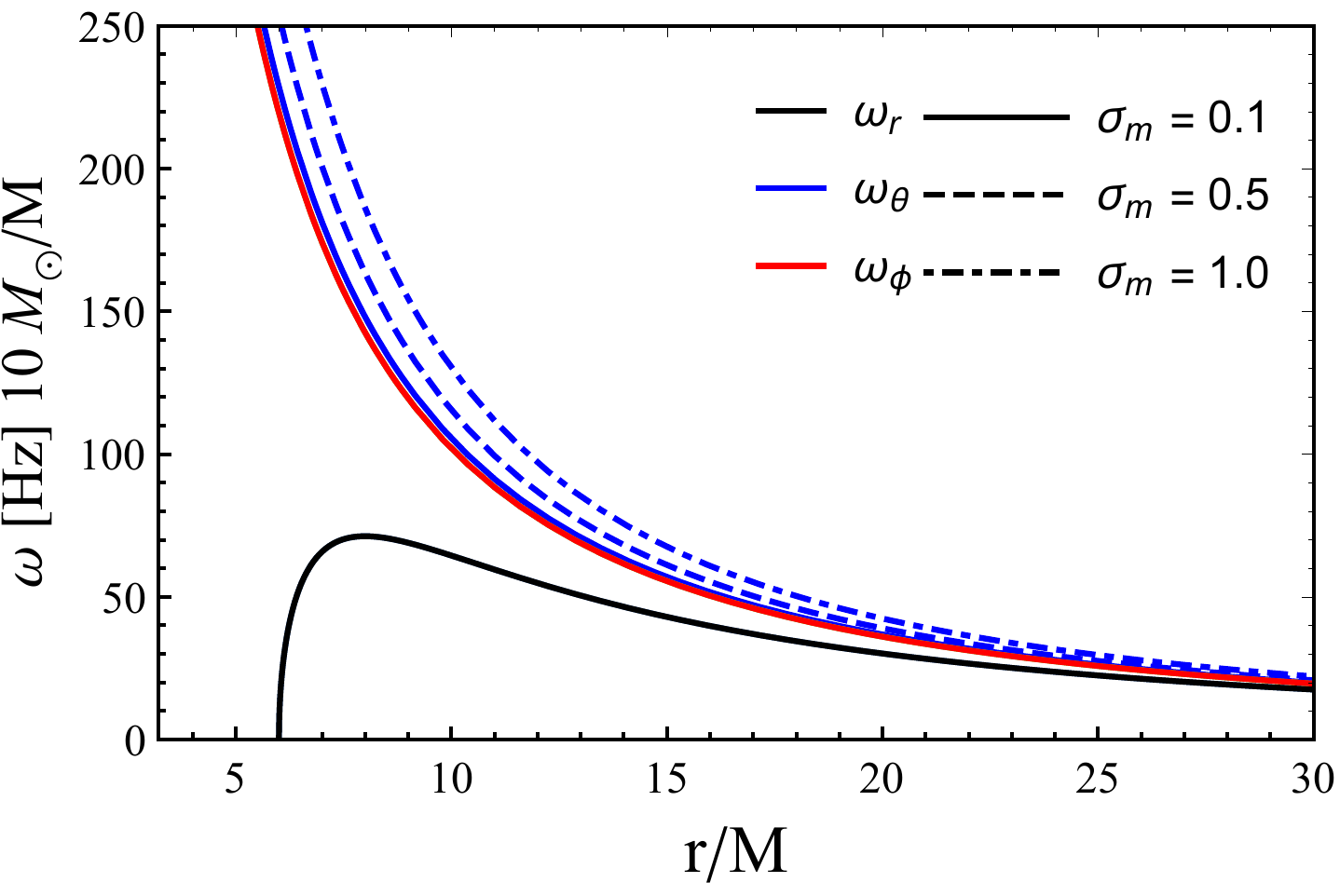}
   \end{tabular}
\caption{\label{fig:qpo_freq_charge} The radial profile of the epicyclic frequencies $\omega_r$, $\omega_{\theta}$ and $\omega_{\phi}$ as a function of $r/M$ for a charged particle case. The first two panels: $\omega_r$, $\omega_{\theta}$ and $\omega_{\phi}$ are plotted for various combinations of the charge coupling parameter $\pm g_m$ while keeping fixed $\sigma=0$. The right panel: $\omega_r$, $\omega_{\theta}$ and $\omega_{\phi}$ are plotted for various combinations of the charge coupling parameter $\sigma_m$ while keeping fixed $g_m=0$. Note that we set $Q_m=0.1$ for all plots. }
\end{figure*}

In this section, we study the frequencies of epicyclic motion of test particle moving in stable circular orbits. It is well known that orbits that occur between $r_{ph} < r < r_{ISCO}$ are unstable, and consequently stable orbits occur above the innermost stable circular orbit, i.e., $r > r_{ISCO}$ that is defined by the effective potential's minimum. Note that in case there are small changes in $r = r_0 + \delta r$ and $\theta=\pi/2+\delta \theta$, as a result, epicyclic motion appears around equilibrium, i.e., the particle oscillates with radial and latitudinal frequencies as stated by a liner harmonic oscillator that is written in terms of small perturbations $\delta r$ and $\delta \theta$ as 
\begin{eqnarray}
\label{Eq:har_osc}
\delta\ddot{ r}&+&\bar{\Omega}_r^2 \delta r = 0\ , \\ 
\delta\ddot{ \theta}&+&\bar{\Omega}_\theta^2 \delta\theta = 0\ ,
\end{eqnarray}
where $\bar{\Omega}_r$ and $\bar{\Omega}_\theta$, respectively, refer to the radial and the latitudinal  frequencies of epicyclic motion. Hereafter we need to define the radial and the latitudinal  frequencies for epicyclic oscillation, measured by a local observer. For that we recall Eq.~(\ref{Eq:sepham}) to obtain these frequencies that are given by~\cite{Shaymatov20egb,Stuchlik21_qpo,Shaymatov22c}
\begin{eqnarray}
\label{baromegas}
\bar{\Omega}_r^2 &=& \frac{1}{g_{rr}}\frac{\partial^2 H_{\textrm{pot}}}{\partial r^2}\ ,\\ 
\bar{\Omega}_\theta^2 &=& \frac{1}{g_{\theta\theta}}\frac{\partial^2 H_{\textrm{pot}}}{\partial \theta^2}\ , \\ 
\bar{\Omega}_\phi &=& \frac{1}{g_{\phi\phi}}\Big(\mathcal{L}-\frac{q}{m} A_\phi\Big)\, .
\end{eqnarray}
\begin{figure*}
\begin{tabular}{c c c}
  \includegraphics[scale=0.4]{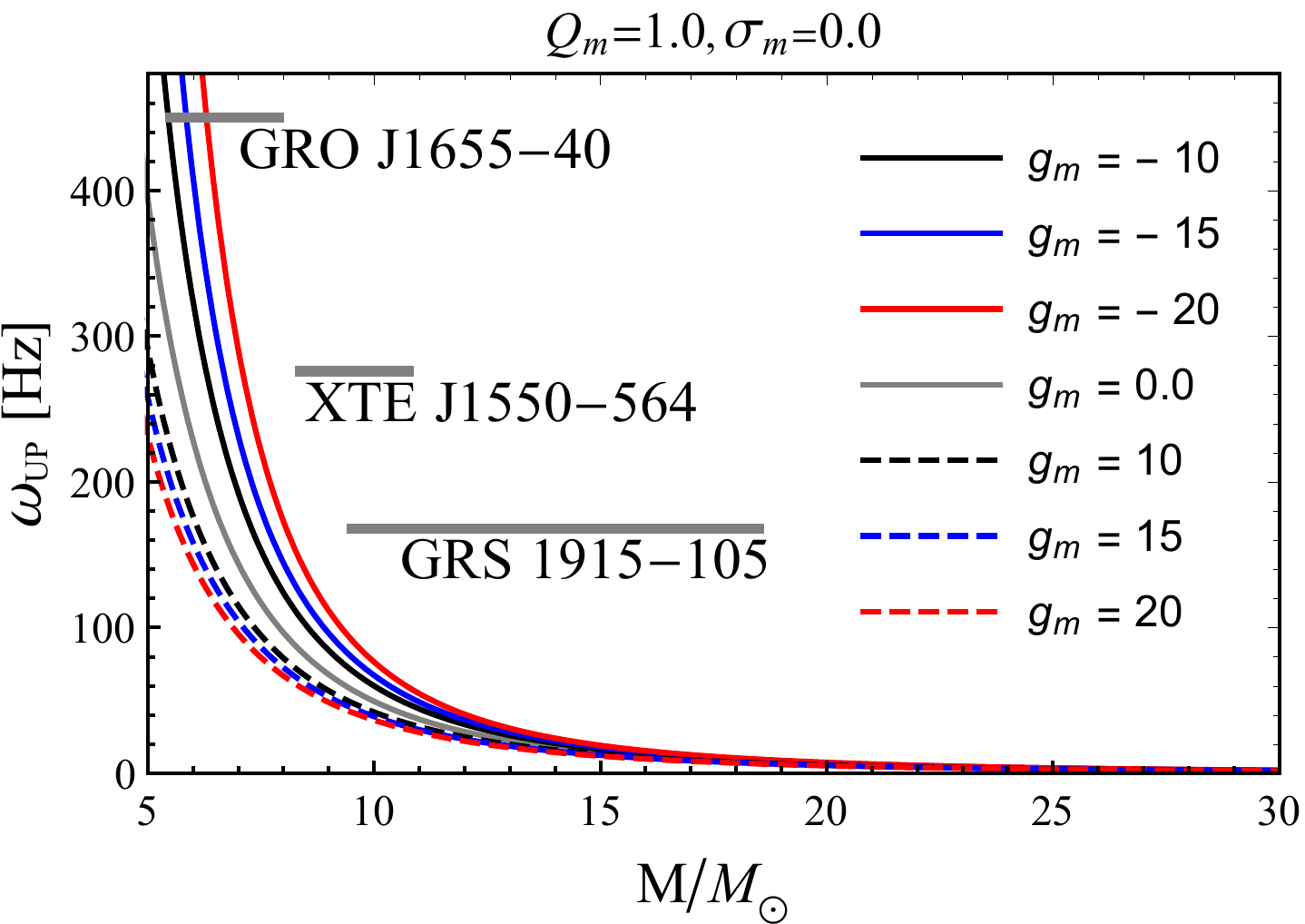}\hspace{-0.0cm}
   &  \includegraphics[scale=0.4]{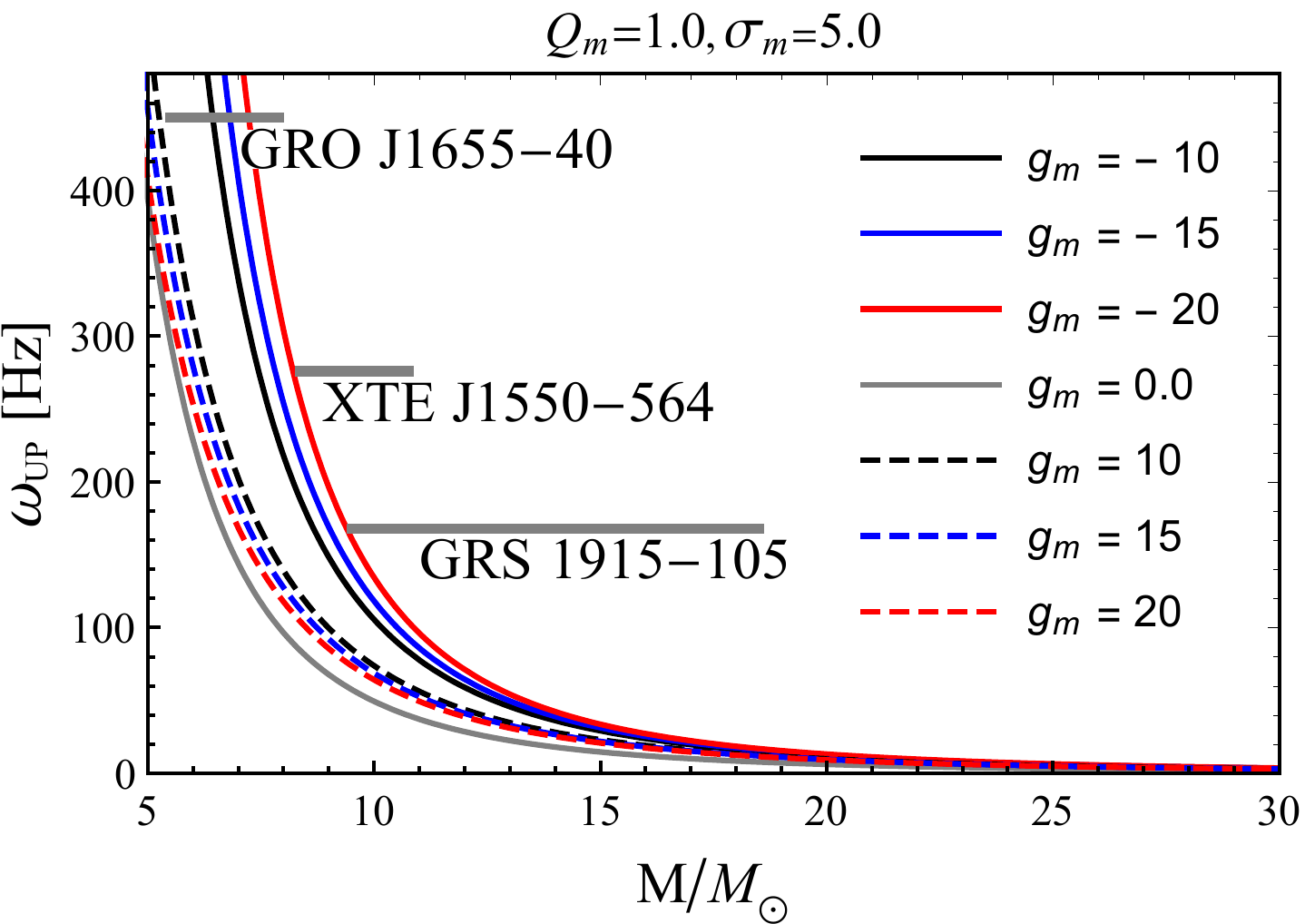}
   \includegraphics[scale=0.4]{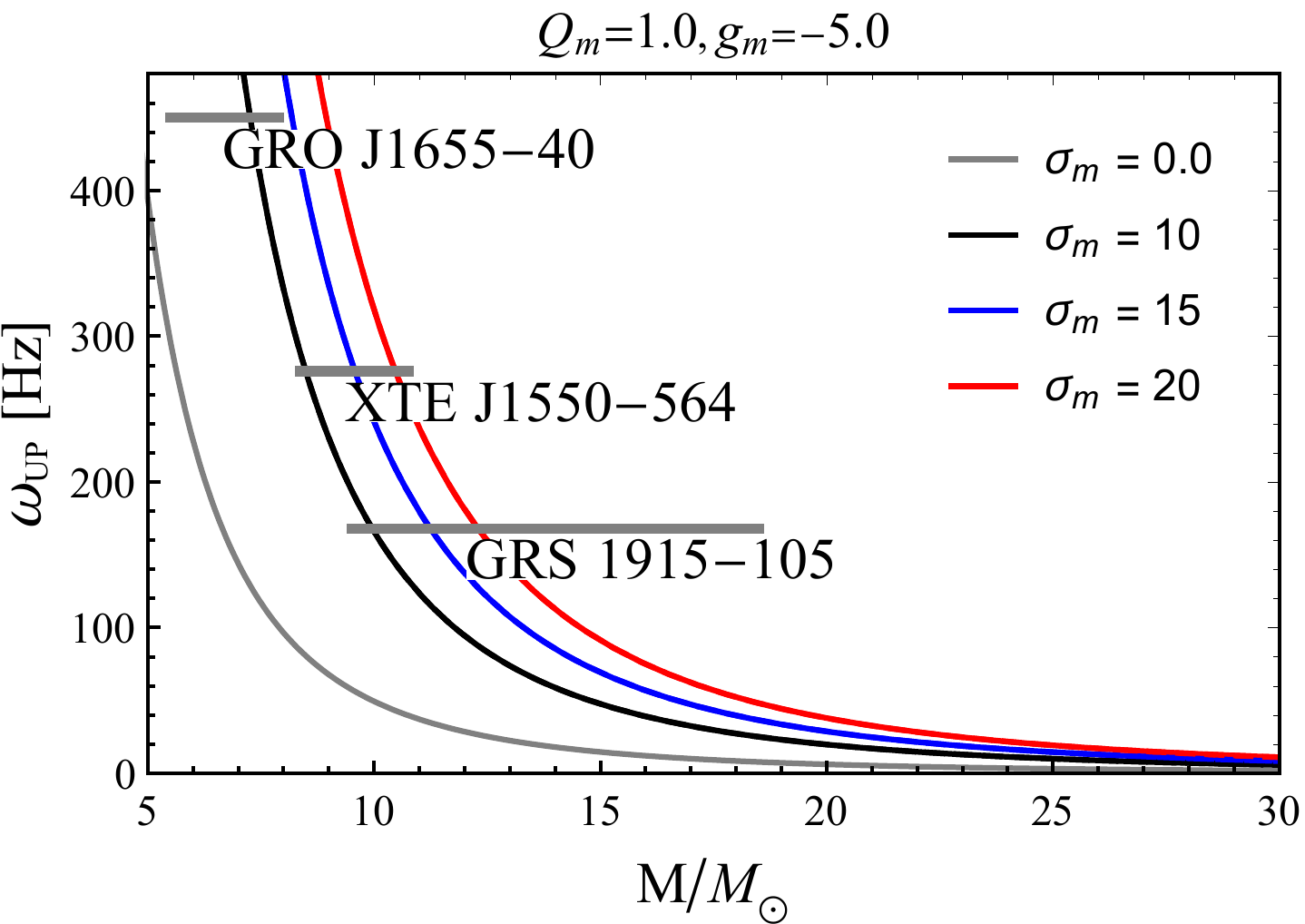}
   \end{tabular}
\caption{\label{fig:masslimit} The mass relation profile of upper frequency for various combinations of the coupling parameters $g_m$ and $\sigma_m$. The left panel:  $\omega_{UP}$ is plotted for various combinations of positive values of $g_m$ in the case of fixed $Q_m=1.0$ and $\sigma=0$. The middle panel: $\omega_{UP}$ is plotted for various combinations of negative values of $g_m$ in the case of fixed $Q_m=1.0$ and $\sigma=5$. The right panel:  $\omega_{UP}$ is plotted for various combinations of values of $\sigma_m$ while keeping $Q_m=1.0$ and $g_m=5$ fixed. Note that the resonance frequency ratio has been set  $\omega_r:\omega_{\theta}=3:2$ for high frequency QPOs.}
\end{figure*}
As was discussed above, to occur the periodic motion the particle needs to move in stable circular orbits around a black hole along with the fundamental frequencies, as well as the specific energy and angular momentum. Note that we consider $u^{\alpha}= (u^{t}, 0, 0,u^{\phi})$ for the periodic motion at the circular orbits. Taking normalization condition $u^{\alpha}u_{\alpha}=-1$ into account we write the following equations 
\begin{eqnarray}\label{Eq:ut}
u^{t}&=&\frac{1}{\sqrt{-g_{tt}-\Omega^2g_{\phi\phi}}}\, ,\\
\label{Eq:en}
\mathcal{E}&=&-\frac{g_{tt}}{\sqrt{-g_{tt}-\Omega^2g_{\phi\phi}}}+\frac{g_m}{r}\, ,\\
\label{Eq:ang}
\mathcal{L}&=&\frac{g_{\phi\phi}\Omega}{\sqrt{-g_{tt}-\Omega^2g_{\phi\phi}}}-{\sigma_m}\cos\theta\, ,
\end{eqnarray}
where we have defined $\Omega= \frac{d\phi}{dt}$ referred to as the so-called Keplerian frequency given in Eq~(\ref{Eq:kep}). Note that we have analysed $\Omega$ in the previous section; see Table \ref{tab:table1} and Fig.~\ref{fig:freq_vel}. However, we further apply it in analysing both the radial and the latitudinal frequencies of epicyclic motion. 
Taking Eqs.~(\ref{Eq:ut}-\ref{Eq:ang}) and (\ref{Eq:kep}) altogether we derive $\Omega_{r}$ and $\Omega_{\theta}$ as  
\begin{eqnarray}\label{Eq:wr}
\bar{\Omega}_{r}^2&=& \frac{Q_m^2-M r}{M^2 r^6 \Big(6 M^2 r-2 M \left(2 Q_m^2+r^2\right)+Q_m^2 r\Big)}\nonumber\\&\times&
\Bigg[g_m^2 \left(Q_m^2-M r\right) \Big(6 M^2 r-2 M \left(2 Q_m^2+r^2\right)\nonumber\\&+& Q_m^2 r\Big)+2 \sqrt{2}\, g_m r \left(Q_m^2-M r\right) \Big(M \left(2 Q_m^2+r^2\right)\nonumber\\&-&2 Q_m^2 r\Big) \sqrt{\frac{M \Big(4 M Q_m^2+2 M r^2-r \left(6 M^2+Q_m^2\right)\Big)}{\left(Q_m^2-M r\right)^2}}\nonumber\\&-&M r^3 \left(2 M^2-Q_m^2\right) \left(r-6 M\right)\Bigg]\, ,\\ \nonumber\\
\label{Eq:wth}
\bar{\Omega}_{\theta}^2&=& \frac{1}{M r^4 \Big(6 M^2 r-2 M \left(2 Q_m^2+r^2\right)+Q_m^2 r\Big)}\nonumber\\&\times &\Bigg[M\sigma_m^2 \Big(6 M^2 r-2 M \left(2 Q_m^2+r^2\right)+Q_m^2 r\Big)\nonumber\\&+& M r^3 \left(Q_m^2-2 M^2\right)\Bigg]\, .
\end{eqnarray}
We wish to analyse the above epicyclic frequencies.  As mentioned previously the frequencies above can be measured by a local observer; however, they need  to be measured by distant observers far away from a black hole vicinity at spatial infinity. Hence, we need to transform $\Omega_i$ into the new definition of $\omega_i$ so that it would be possible for a distant observer to measure them from spatial infinity. For that we introduce the redshift factor $\bar{\Omega}\rightarrow{-\bar{\Omega}}/{\left(g^{tt}\mathcal{E}\right)}$, transformed from the proper time $\tau$ to the time $t$ measured at spatial infinity~\cite{Shaymatov20egb,Stuchlik21_qpo,Shaymatov22c}. Consequently, the relation between these two $\bar{\Omega_i}$ and $\omega_{i}$ frequencies can be written as follows with $G$ and $c$
\begin{eqnarray}\label{Eq:rel}
\omega=\frac{1}{2\pi}\frac{c^3}{GM}\frac{\bar{\Omega}}{(-g^{tt})\mathcal{E}}\, . 
\end{eqnarray}
One can then state that the observer located at spatial infinity would be capable of measuring these epicyclic frequencies directly and analysing their nature for the primary source of the QPOs allowing to obtain information about the accretion discs around supermassive black holes, as well as the nature of the geometry.

\begin{figure*}
    \centering
    \includegraphics[scale=0.45]{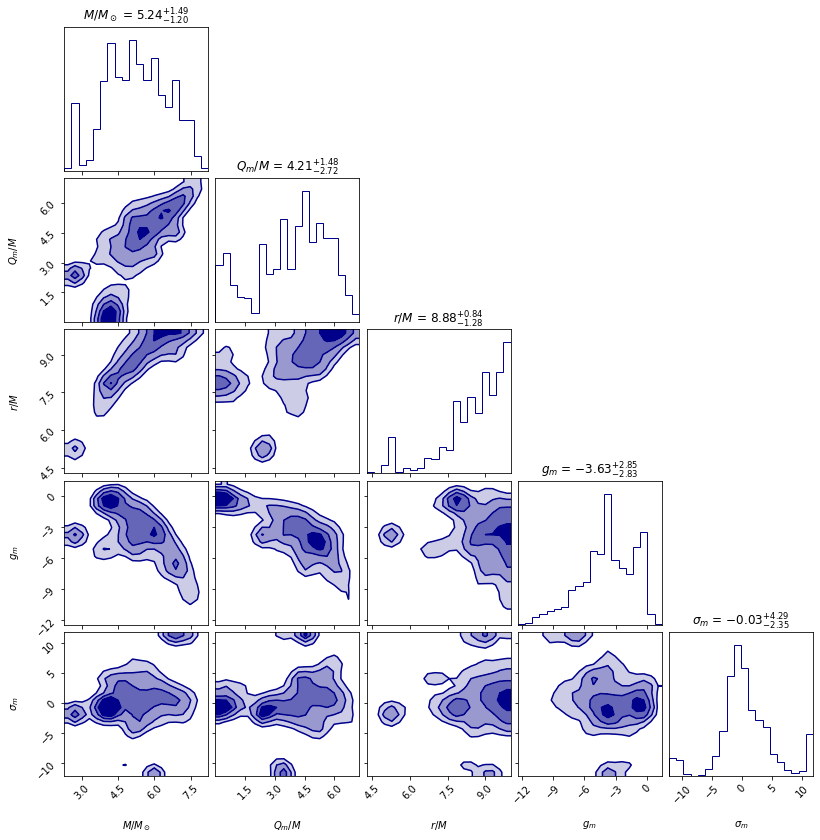} 
    \caption{\label{fig:con1} The contours plots and the best-fit
parameters within $1-\sigma$ confidence region for the microquasar GRO J1655-40}
    \label{temperature1}
\end{figure*}

\begin{figure*}
    \centering
    \includegraphics[scale=0.45]{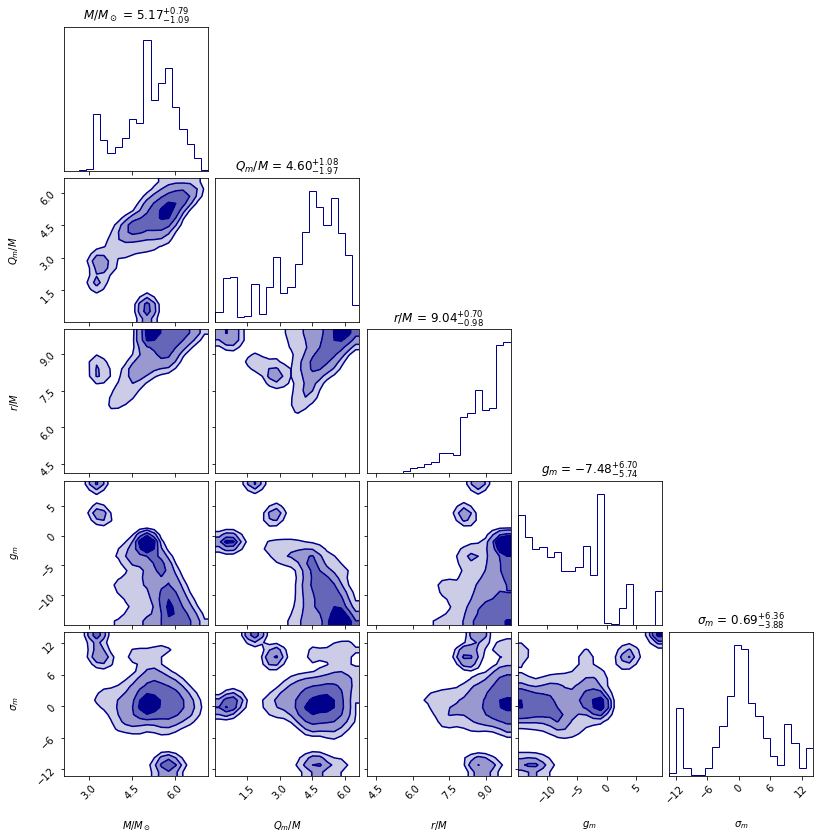} 
    \caption{\label{fig:con2} The contours plots and the best-fit
parameters within $1-\sigma$ confidence region for the microquasar XTE J1550-564}
    \label{temperature1}
\end{figure*}

\begin{figure*}
    \centering
    \includegraphics[scale=0.45]{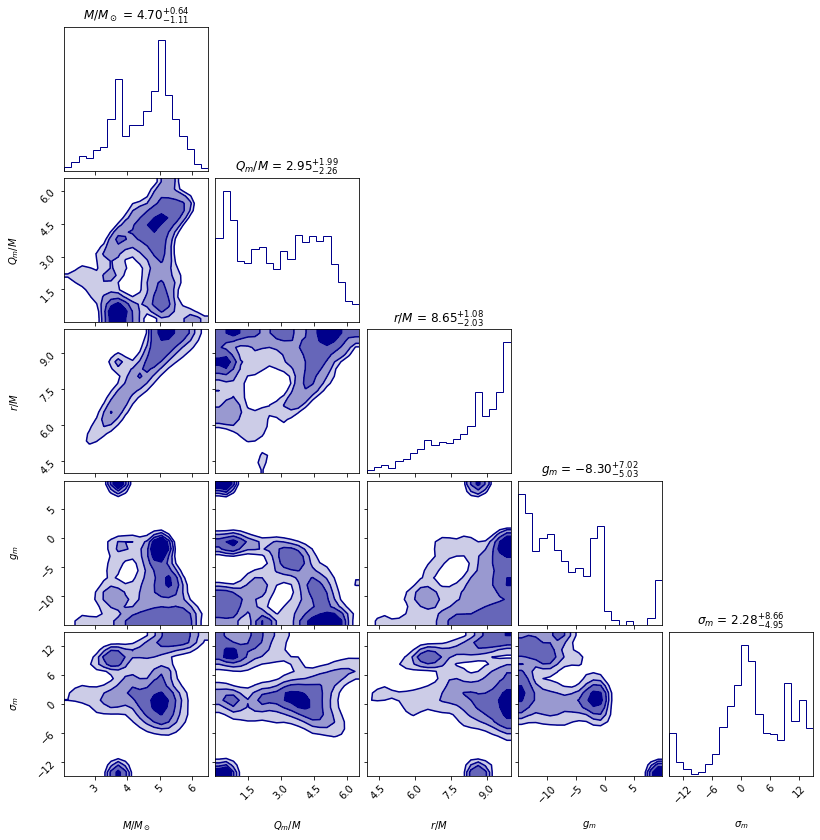} 
    \caption{\label{fig:con3} The contours plots and the best-fit
parameters within $1-\sigma$ confidence region for the microquasar GRS 1915+105}
    \label{temperature1}
\end{figure*}

Let us then further analyse the behaviour of above mentioned epicyclic frequencies that would be measured by the distant observer located at spatial infinity. In Figs.~\ref{fig:qpo_freq} and \ref{fig:qpo_freq_charge} we demonstrate the radial profile of these three frequencies for both neutral and electrically and magnetically charged particles moving in the stable circular orbits around the magnetically charged stringy black hole for various combinations. Fig.~\ref{fig:qpo_freq} reflects the role of black hole magnetic charge on the radial profile of epicyclic frequencies for a neutral particle case. From Fig.~\ref{fig:qpo_freq} the radial and latitudinal frequencies for the distant observer decrease as $Q_m$ increases, thus allowing their shape to shift down to smaller $\omega$, while the orbital frequency $\omega_{\phi}$ remains unchanged. Similarly, Fig.~\ref{fig:qpo_freq_charge} reflects the impact of the charge coupling parameters on the radial profile of the epicyclic frequencies. As can be seen from the left and middle of panels Fig.~\ref{fig:qpo_freq_charge} the behaviour of the radial frequency strongly gets impacted by the coupling parameter$g_m$ for both negative and positive cases, i.e., it respectively increases and decreases with increasing the coupling parameter $\pm g_m$ in the case of vanishing $\sigma=0$. We notice that $\omega_{\theta}$ and $\omega_{\phi}$ slightly get affected as a result of increasing the coupling parameter $\pm g_m$, i.e., both significantly grow and decrease on a small scale as compared to the one for the radial frequency $\omega_r$, as seen in the first two panels of Fig.~\ref{fig:qpo_freq_charge}. We interestingly observe that the coupling parameter $\sigma_m$ can only affect the behaviour of the latitudinal frequency $\omega_{\theta}$ in the case of vanishing $g_m$ while keeping $Q_m$ fixed. Unlike the latitudinal frequency, the radial and orbital frequencies remain unchanged; see the right panel of Fig.~\ref{fig:qpo_freq_charge}.   \begin{figure*}
\begin{tabular}{c c }
  \includegraphics[scale=0.4]{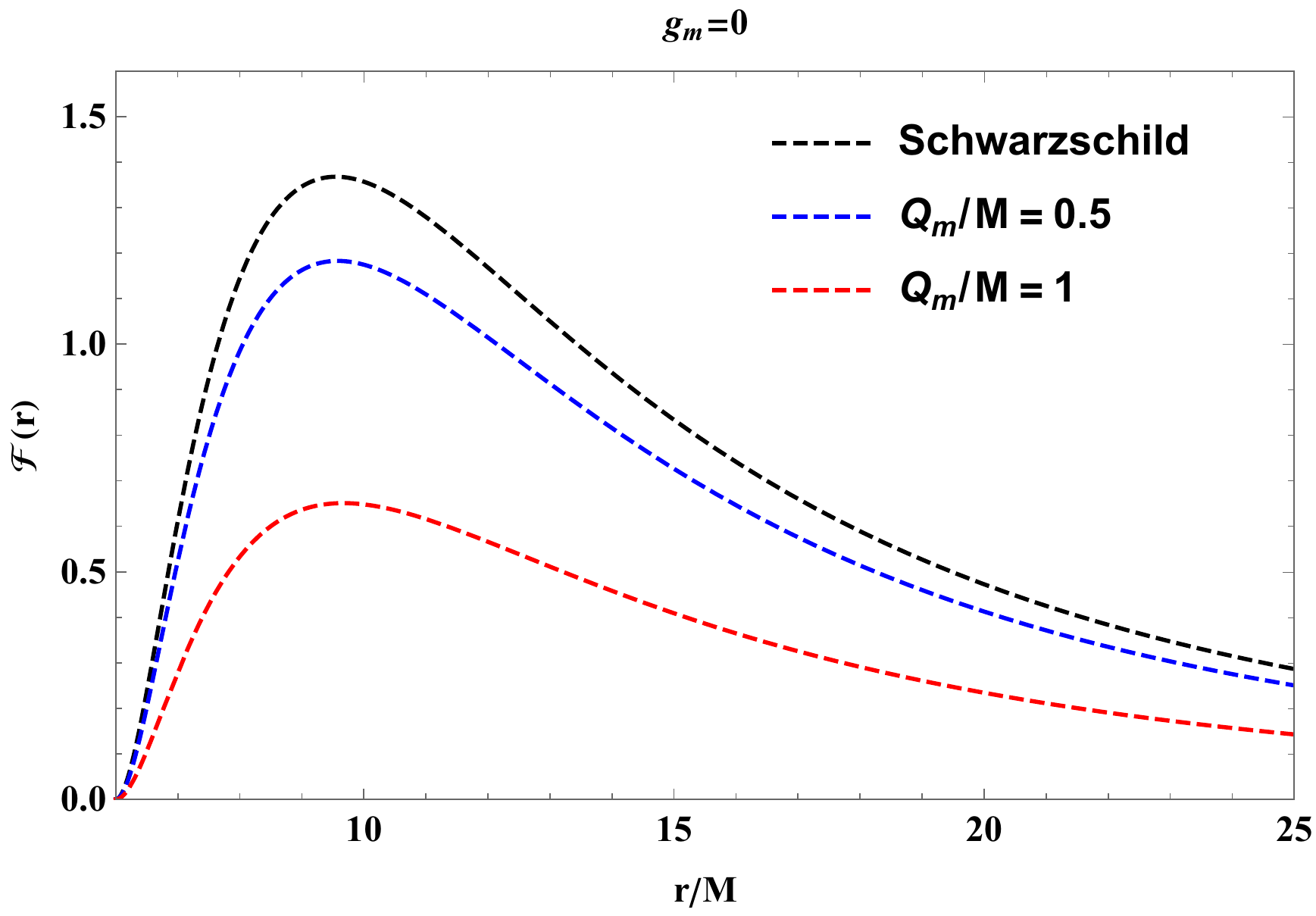}\hspace{-0.0cm}
   &  \includegraphics[scale=0.4]{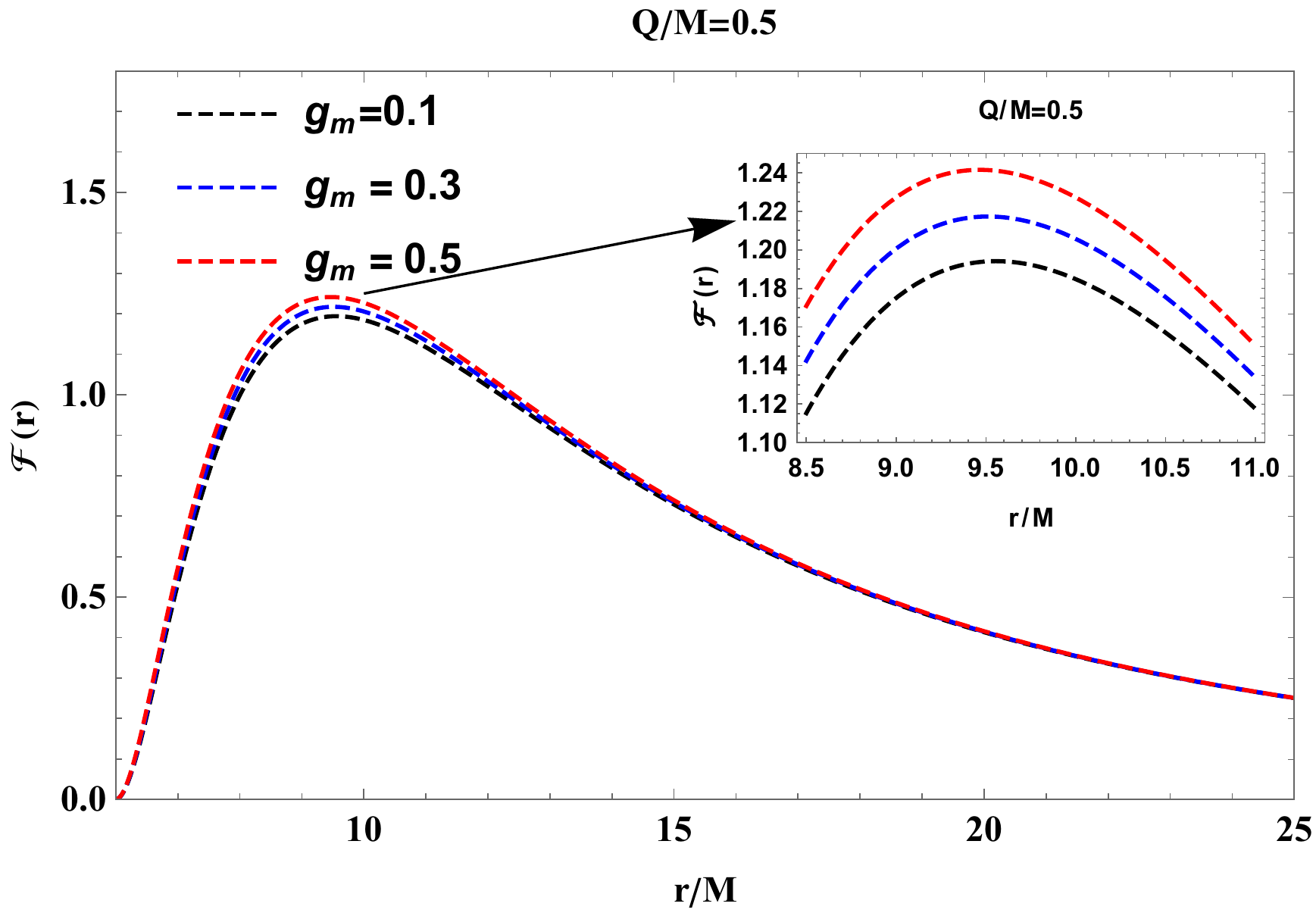}
   \end{tabular}
   \begin{tabular}{c c }
   \includegraphics[scale=0.4]{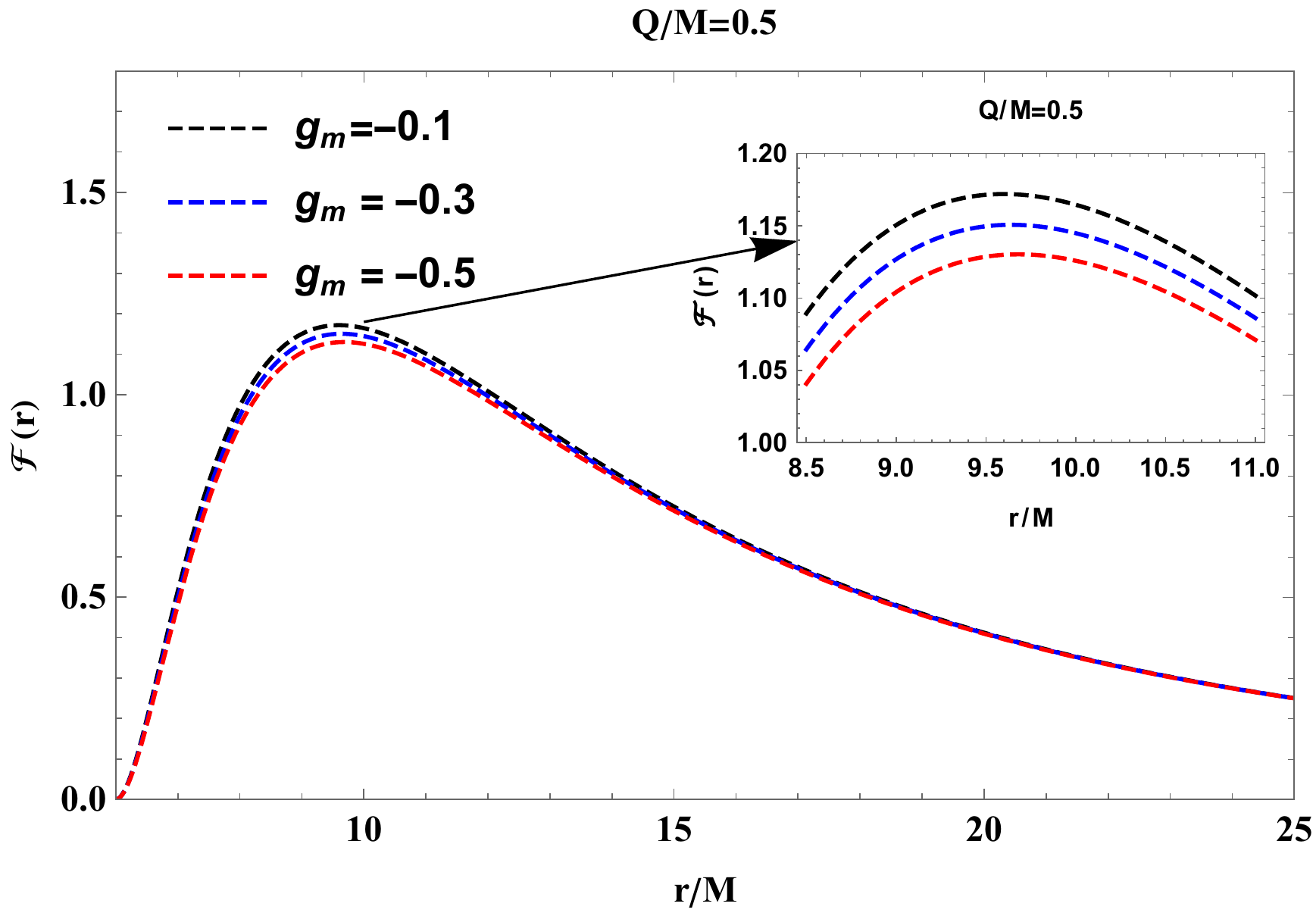}
   &  \includegraphics[scale=0.4]{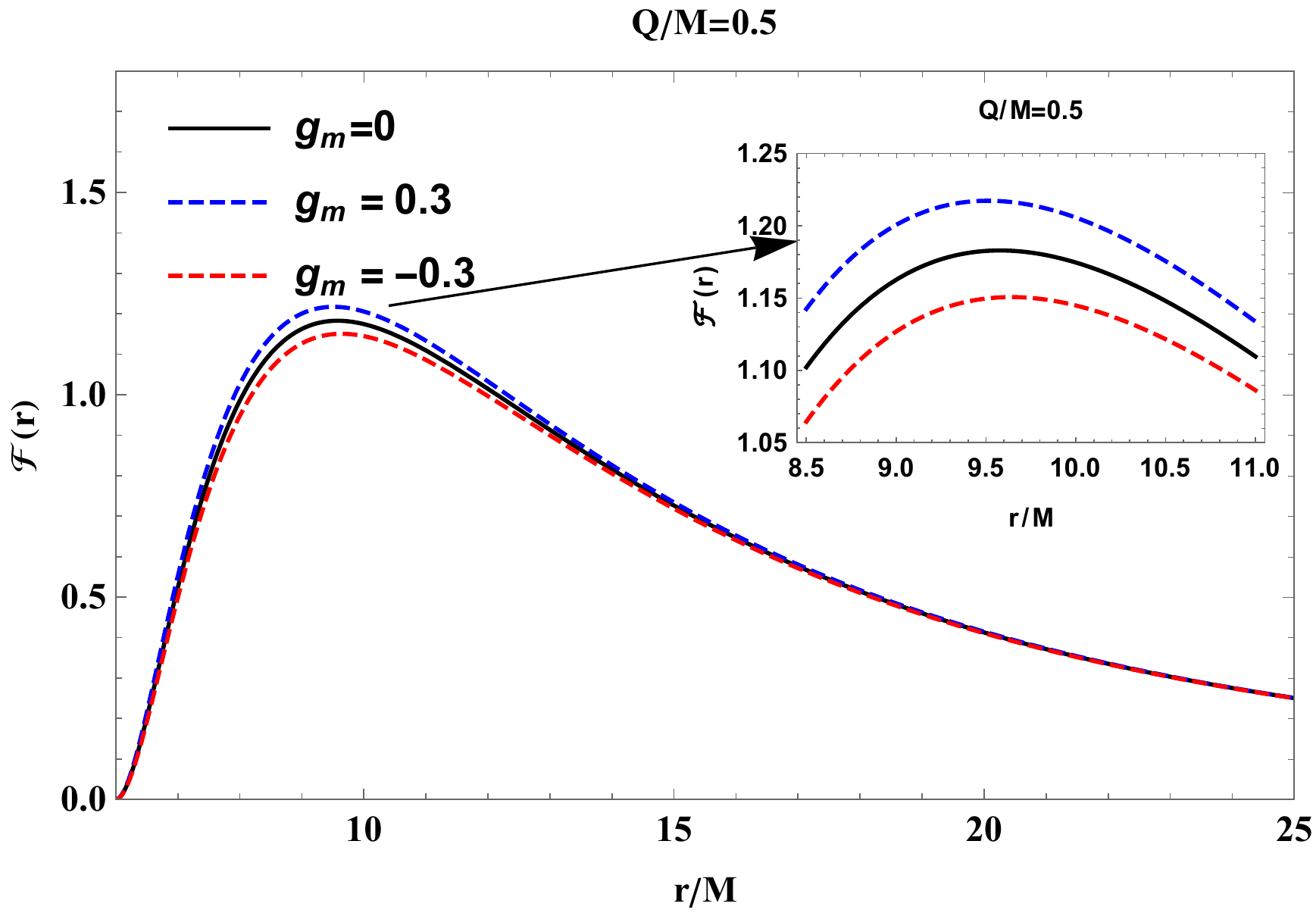}
   \end{tabular}
\caption{\label{fig:flux} Radial dependence of the flux of electromagnetic radiation of the disk on different values of  $Q_m$ and $g_m$ parameters. Numerical evaluation of the flux $\mathcal{F}$ divided by $10^{-5}$ of the accretion disk as a function of $r/M$}
\end{figure*}

 We note that the quasi-periodic oscillations can be expected to be very potent tests in revealing not only theoretical models but also unknown aspects related to the precise measurements of black hole parameters. With this in view, the high-frequency quasi-periodic oscillations (HF QPO), the X-ray power observed in microquasars regarded as a primary source of the low-mass X-ray binary systems consisting of a black hole or a neutron star, become increasingly  important and come into play to have important astrophysical applications. In the strong field regime, the particles generally oscillate radially and vertically with the corresponding frequencies near the ISCO radius, thereby resulting in yielding the quasiperiodic power spectra observed by a distant observer. Hence, these mentioned HF QPOs can be respectively described by upper and lower frequencies with the fixed ratio $\omega_{U}:\omega_{L}=3:2$ in most cases \cite{Abramowicz:2003xy,Horak:2006sw,Torok05A&A,Remillard06ApJ}. Here we consider three HF QPOs sources (i.e., $\mbox{GRO 1655-40}$, $\mbox{XTE 1550-564}$ and $\mbox{GRS 1915 + 105}$) observed with upper and lower frequencies that can be identified explicitly by $\omega_U=\omega_r$ and $\omega_L=\omega_{\theta}$. Irrespective of the fact that the occurrence of $\omega_U:\omega_L$ = 3:2 has been observed for these three microquasars for twin HF QPOs there would be, of course, other specific models with various resonances. We do however restrict ourselves to the three microquasars for the parameters here considered to best fit $\omega_U:\omega_L$ = 3:2. In Fig.~\ref{fig:masslimit} we show the mass relation profile of upper frequency with respect to astrophysical data for various combinations of the charge coupling parameters $g_m$ and $\sigma_m$. As can be seen from the left panel of Fig.~\ref{fig:masslimit}, the upper frequency respectively decreases and increases as a consequence of an increase in the value of charge coupling parameter $\pm g_m$ with respect to the neutral particle case. However, the effect of $g_m$ does not well fit the observed data yet. For the best fit of observed data, the combined effect of the charge coupling parameters $g_m$ and $\sigma_m$ becomes increasingly important, as seen in the middle and right panels of Fig.~\ref{fig:masslimit}. The presence of $\sigma$ reinforces the effect that causes the shape of the upper frequency to shift towards the right in order to best fit the observed data. Hence, the best fits correspond to the combined effect of the charge coupling parameters, especially as a consequence of increasing $\sigma_m$.

 \section{Observational Constraints}\label{Sec:Constrain}

In this section, we would like to put observational constraints on the parameters of the stringy black hole as well as the coupling parameters. Toward this goal, we will use the experimental data for three well known microquasars. As we have already mentioned in the previous section, the appearance of two peaks in the X-ray power density spectra of Galactic microquasars has attracted a lot of interest in order to explain the value of the 3:2-ratio from a theoretical point of view \cite{Strohmayer:2001yn}. Specifically, we will use the occurrence of $\omega_U:\omega_L=3:2$ with lower $\omega_L$ Hz QPO and upper $\omega_U$ Hz QPO observed for the following three microquasars:  GRO J1655-40, XTE J1550-564 and GRS 1915+105.

The corresponding frequencies of these microquasars are given by: \cite{Strohmayer:2001yn}
\begin{multline}\label{pr1}
\text{GRO J1655-40 : }
\omega_U=450\pm 3 \text{ Hz},\;\omega_L=300\pm 5 \text{ Hz},
\end{multline}
\begin{multline}\label{pr2}
\text{XTE J1550-564 : }
\omega_U=276\pm 3 \text{ Hz},\;\omega_L=184\pm 5 \text{ Hz},
\end{multline}
\begin{multline}\label{pr3}
\text{GRS 1915+105 : }
\omega_U=168\pm 3 \text{ Hz},\;\omega_L=113\pm 5 \text{ Hz},
\end{multline}
%
The precise nature behind QPOs is not yet well understood. However, there are different explanations for the twin values of the QPOs. For example, one possible explanation of QPOs is known as the phenomenon of resonance. According to this view,  near the vicinity of the ISCO the in-falling particles can perform radial as well as vertical oscillations and, in general, the two oscillations couple non-linearly yielding the observed quasiperiodic power spectra \cite{Abramowicz:2003xy,Horak:2006sw}. In the present study, besides the mass, we are interested and would like to understand more about the role of magnetic charge of the central black hole on QPOs. There are, however, different models representing different types of resonances. In the current work, we shall assume the parametric resonance that is given by 
\begin{equation}\label{as1}
\omega_U = \omega_r \mbox{\, \, and\, \,}  \omega_L =\omega_{\theta}\ .
\end{equation}
Next, in order to constrain the parameters that characterize the black hole spacetime we will assume a five parameter model for the QPO frequency and we will perform a Monte Carlo simulation with a $\chi$-square analysis which is given by 
\begin{equation}
\chi^{2}(M,Q_m,r,g_m,\sigma_m)=\frac{(\omega_{r}-\omega_{1\mathrm{U}})^{2}
}{\sigma^2_{1\mathrm{U}}}\\\notag
+\frac{(\omega_{\theta}-\omega_{1\mathrm{L}})^{2}%
}{\sigma^2_{1\mathrm{L}}}\ .
\end{equation}
In addition, we will use a uniform prior for parameters given by $M/\textup{M}_\odot \in [2,12]$, $Q_m/M \in [0,8]$,  $r/M\in [4,10]$, $g_m \in [-15,15]$, and $\sigma_m \in [-15,15]$, respectively. \\

Below, we present our findings for each case separately:
\begin{itemize}
    \item  Microquasar GRO J1655-40
\end{itemize}
 For this case, within 1$\sigma$ we have obtained the following best fit parameters: black hole mass $M/\textup{M}_\odot=5.24^{+1.49}_{-1.20}$, magnetic charge $Q_m/M=4.21^{+1.48}_{-2.72}$, along with $r/M=8.88^{+0.84}_{-1.28}$. For the coupling parameters $g_m$ and $\sigma_m$ we obtained: $g_m=-7.48^{+6.70}_{-5.74}$ and $\sigma_m=0.69^{+6.36}_{-3.88}$, respectively. The parametric plots for this case are presented in Fig.~\ref{fig:con1}.
 \begin{itemize}
\item Microquasar XTE J1550-564:
\end{itemize}
 For this microquasar within 1$\sigma$ confidence we obtained: for the black hole mass $M/\textup{M}_\odot=5.17^{+0.79}_{-1.09}$, for the magnetic charge $Q_m/M=4.60^{+1.08}_{-1.97}$ and $r/M=9.04^{+0.70}_{-0.98}$. For the coupling parameters we have obtained $g_m=-7.48^{+6.70}_{-5.74}$ and $\sigma_m=0.69^{+6.36}_{-3.88}$, respectively. Again, for more details, the parametric plots are presented in Fig.~\ref{fig:con2}.

 \begin{itemize}
\item Microquasar GRS 1915+105
\end{itemize}
 Here we find within 1$\sigma$ a black hole mass $M/\textup{M}_\odot=4.70^{+0.64}_{-1.11}$, magnetic charge $Q_m/M=2.95^{+1.99}_{-2.26}$ and $r/M=8.65^{+1.08}_{-2.03}$. For the parameters $g_m$ and $\sigma_m$ we get: $g_m=-8.30^{+7.02}_{-5.03}$ and $\sigma_m=2.28^{+8.66}_{-4.95}$. The parametric plots for this case are presented in Fig.~\ref{fig:con3}.

\section{\label{Sec:accretion}
Accretion disk radiation for magnetically charged stringy black hole}

In this section, we investigate the flux produced by an accretion disk of the magnetically charged stringy black hole. An accretion disk is a disk of gas and dust that orbits around a compact object such as a black hole or a neutron star. As the gas and dust in the disk fall towards the compact object, they heat up and emit radiation in a process known as accretion disk radiation. The accretion disk radiation for a magnetically charged stringy black hole is expected to be influenced by the strong magnetic field of the black hole. The magnetic field can influence the behavior of the gas and dust in the accretion disk, causing it to emit radiation with specific properties.
In particular, the magnetic field can cause the gas and dust in the accretion disk to become highly ionized, meaning that the atoms lose some or all of their electrons. This ionization can lead to the emission of highly energetic radiation, such as X-rays, which can be detected by telescopes.
Additionally, the magnetic field can also cause the gas and dust in the accretion disk to become highly polarized, meaning that the radiation emitted by the disk is preferentially oriented in a specific direction. This polarization can provide clues about the geometry of the magnetic field and the structure of the accretion disk itself.

The accretion disk radiation for a magnetically charged stringy black hole belongs to an area of active research, as it has the potential to provide valuable insights into its properties. 
One can find the flux of the electromagnetic radiation using the following equation ~\cite{Novikov:1973kta,Shakura:1972te,Thorne:1974ve}
\begin{equation}\label{flux}
    \mathcal{F}(r)=-\dfrac{\dot{M_0}}{4 \pi \sqrt{\gamma}}\dfrac{\Omega_{,r}}{(E-\Omega L)^2} \int _{r_{ISCO}}^r (E-\Omega L) L_{,r} d r\ , 
\end{equation}
where $\gamma$ is the determinant of the three-dimensional subspace with coordinates ($t, r , \phi$) and is given by $\sqrt{\gamma}=\sqrt{-g_{tt}g_{rr}g_{\phi \phi}}$. Note that the quantity $\mathcal{F}(r)$ depends on $\dot{M_0}$, which refers to as the disk mass accretion rate and remains unknown. We shall for simplicity choose $\dot{M_0}=1$.  

To find the flux of the electromagnetic radiation we use Eqs.~(\ref{Eq:ut})-(\ref{Eq:ang}). It does however turn out that it is complicated to derive the analytical expression of the flux, and thus we resort it numerically; see Fig. \ref{fig:flux}. As can be seen from the graphs that the flux of the electromagnetic radiation decreases under the effect of the $Q_m$ parameter. We can also see that as the value of $g_m$ increases, so does the flux.  
Moreover, the flux of the black body radiation can be written as $\mathcal{F}(r)=\sigma T^4$ with $\sigma$ as the Stefan-Boltzmann constant. The radial dependence of the disk temperature is represented in Fig.~\ref{temperature1}. This graph depicts the dependence of the disk temperature and energy on the $Q_m$ parameter for fixed values of $g_m$. One can see that the temperature and energy of the disk are increased with the increase in the parameter $Q_m$.

\begin{figure*}
    \centering
    \includegraphics[scale=0.5]{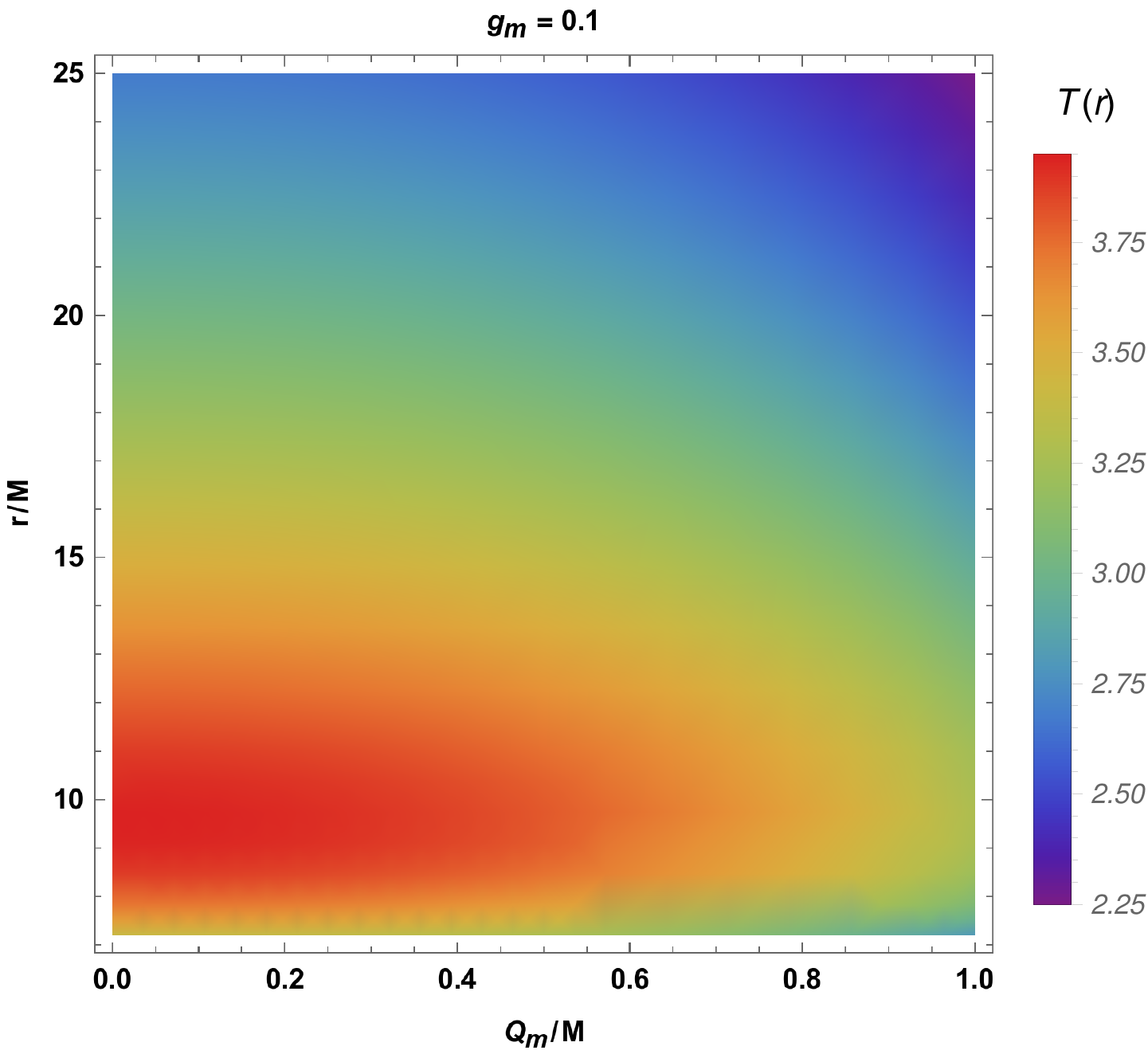}
    \includegraphics[scale=0.5]{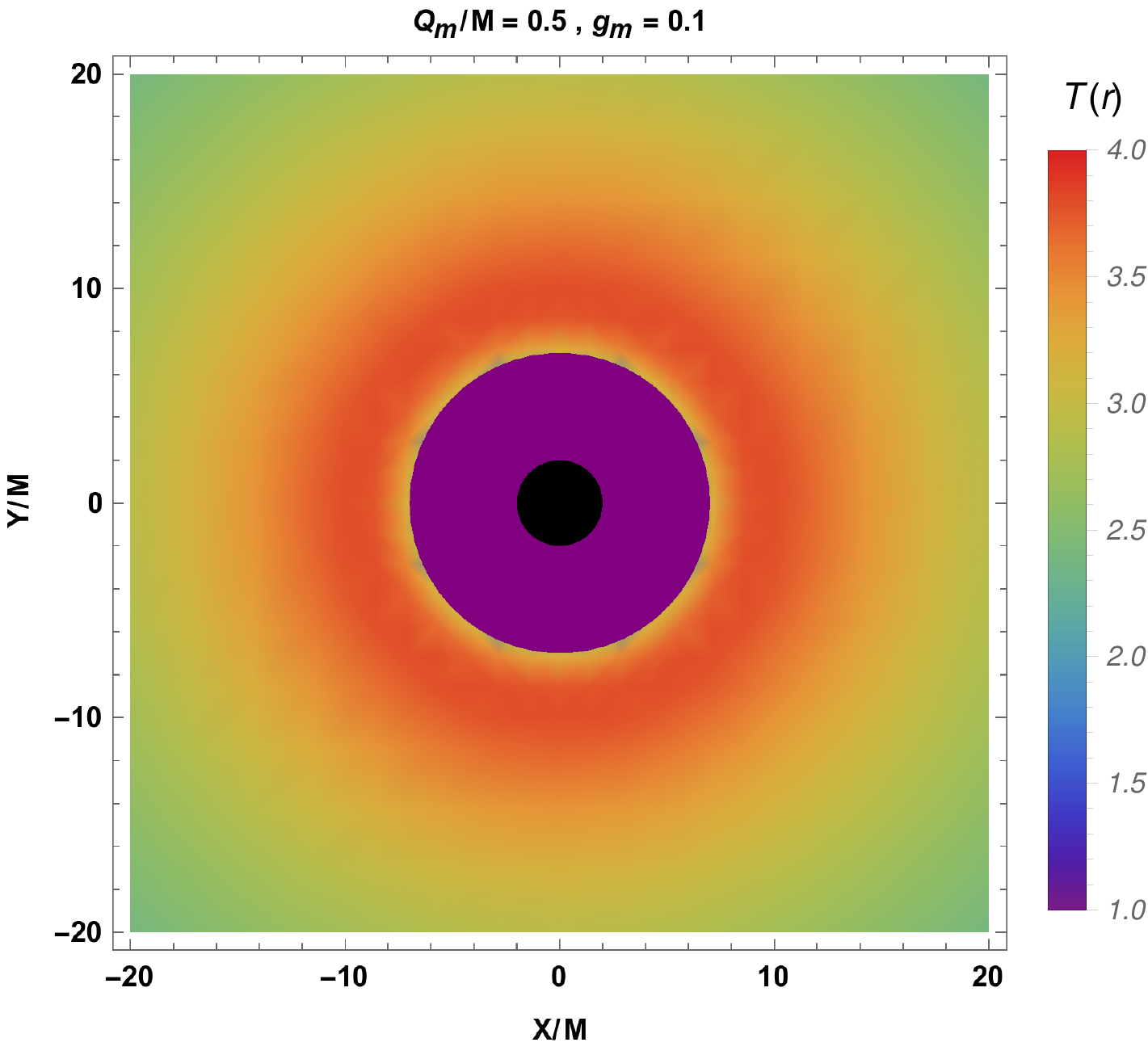}
    
    \caption{Temperature profile of the radial dependence of black hole magnetic charge is plotted in the left panel, while in the right panel temperature profile is plotted on the equatorial $X-Y$ plane (i.e., we
would mean the Cartesian coordinates by $X$ and $Y$) in the form of density plot. }
    \label{temperature2}
\end{figure*}

For being more informative, we demonstrate the temperature graph by "color map" in Fig.~\ref{temperature2}. From Fig.~\ref{temperature2}, one can now clearly notice the dark region inside the accretion disk's inner edge, i.e., the red parts represent the regions that correspond to the disk's maximum temperature.
\begin{figure*}
    \centering
    \includegraphics[scale=0.41]{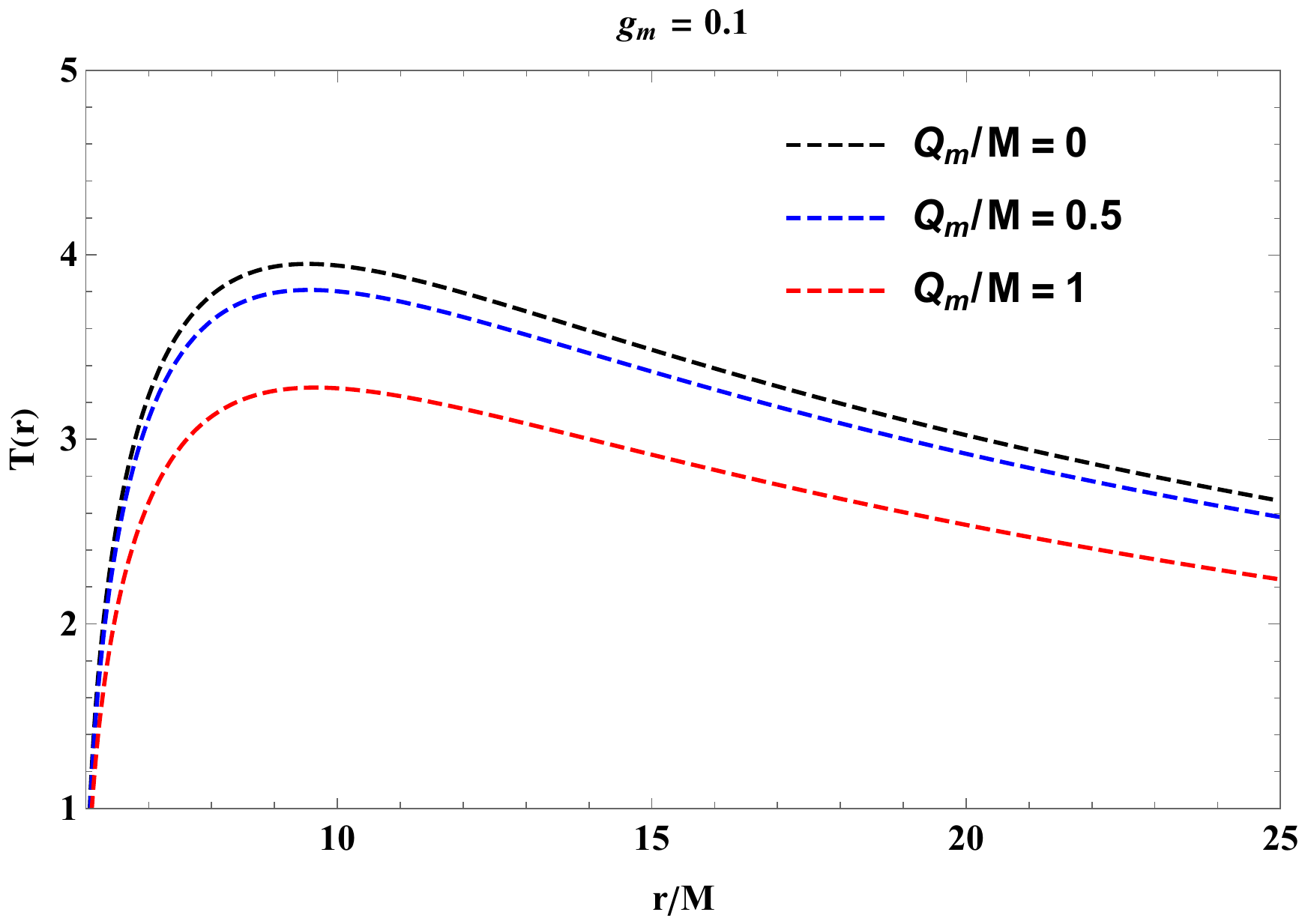}
    \includegraphics[scale=0.45]{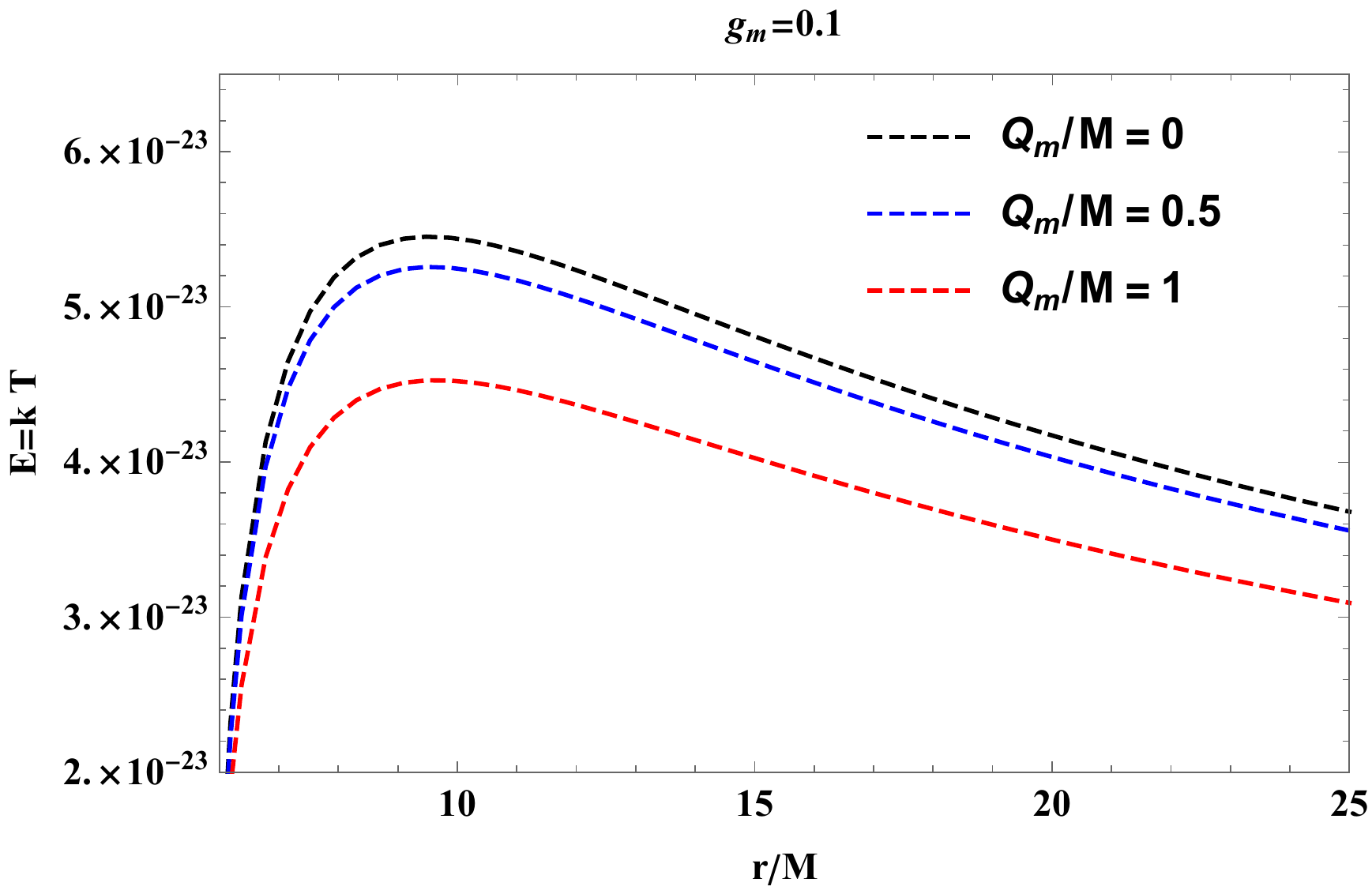}
    
    \caption{The radial dependence of the temperature of the disk (left) and energy (right) for various combinations of $Q_m$ while keeping $g_m$ parameter fixed. }
    \label{temperature1}
\end{figure*}

We then turn to another important quantity referred to as the differential luminosity. It can be written as follows~\cite{Novikov:1973kta,Shakura:1972te,Thorne:1974ve}
\begin{equation}
    \dfrac{d \mathcal{L}_{\infty}}{d \ln{r}}=4 \pi r \sqrt{\gamma} E \mathcal{F}(r)\, .
\end{equation}
\begin{figure}
    \centering
    \includegraphics[scale=0.5]{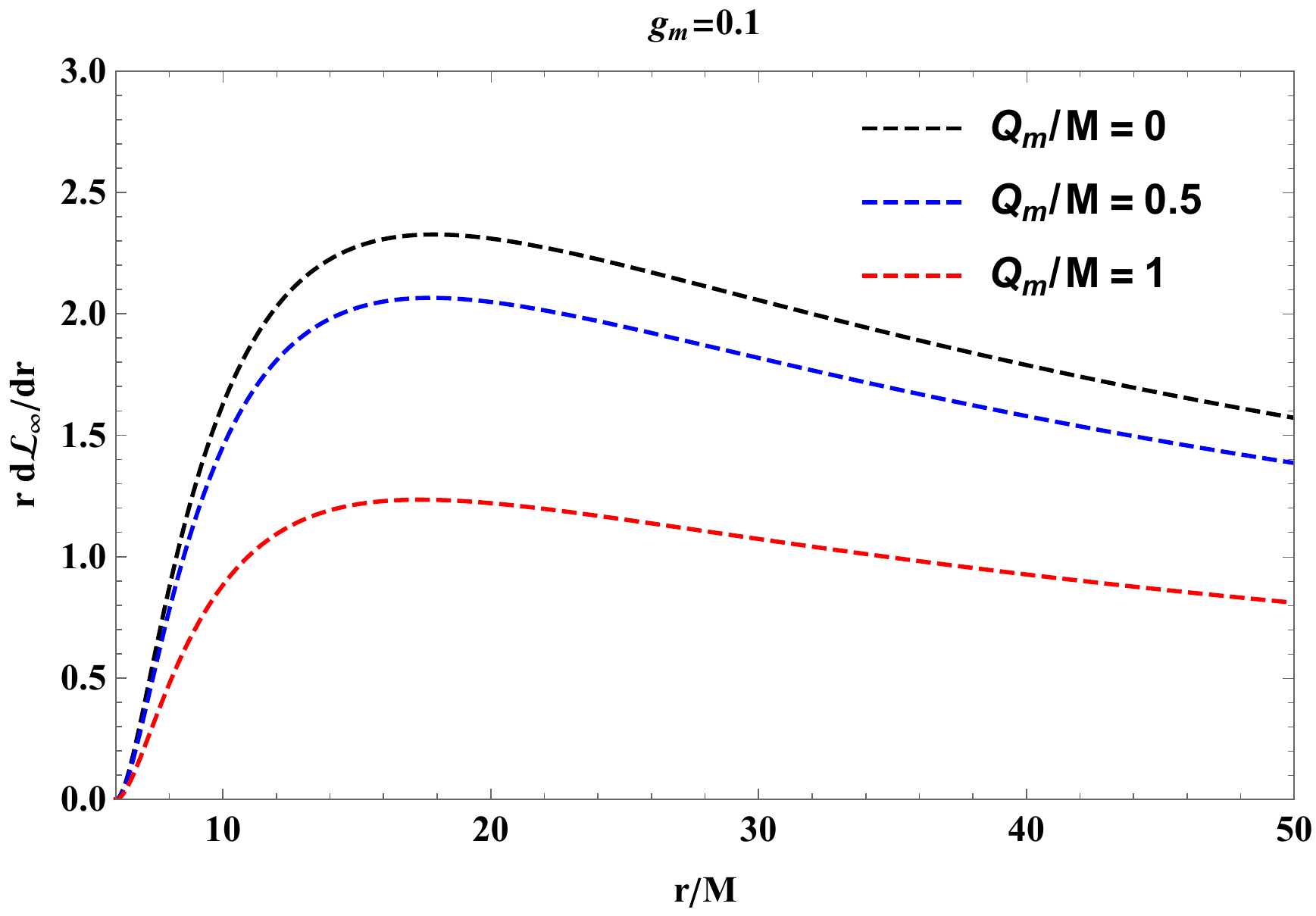}
    \caption{Numerical evaluation of the differential luminosity of the accretion disk scaled in powers of $10^{-2}$ as a function of $r/M$}
    \label{fig:luminosity}
\end{figure}

We assume that the radiation emission can be described by a black body radiation. Bearing this in mind, one can then define the spectral luminosity $\mathcal{L}_{\nu,\infty}$ as a function of the radiation frequency, $\nu$, at infinity~\cite{Boshkayev:2020kle,sym14091765}
\begin{equation}\label{luminosity2}
    \nu \mathcal{L}_{\nu,\infty}=\dfrac{60}{\pi^3} \int_{r_{ISCO}}^{\infty} \dfrac{\sqrt{\gamma} E}{M_T^2}\dfrac{(u^t y)^4}{\exp\Big[{\dfrac{u^t y}{(M_T^2 \mathcal{F})^{1/4}}}\Big]-1}\, ,
\end{equation}
with $y=h \nu /k T_{\star}$, where $h$ and $k$ respectively refer to the Planck constant and the constant of Boltzmann, as well as the total mass $M_T$. Also $T_{\star}$ corresponds to the characteristic temperature and has the relation with the Stefan-Boltzmann law, i.e., $\sigma T_{\star}= \dfrac{\dot{M}_0}{4 \pi M_T^2}$, where $\sigma$ accordingly refer to as the Stefan-Boltzmann constant.

\section{Conclusions}

The presence of an accretion disk is considered to be a powerful mechanism for generating QPOs (quasi-periodic oscillations) in the vicinity of self-gravitating compact objects. As a result, this can yield valuable insights into the spacetime geometry and fields in the strong field regime of the surrounding environment of these objects~\cite{Abramowicz13}. The motion of test particles and their geodesics can be significantly impacted by the presence of existing fields, resulting in changes to observable quantities such as the ISCO (innermost stable circular orbit) and QPO parameters. Electromagnetic fields can serve as a potent means of modifying the geodesic orbits of particles that are in orbit around an accretion disk in the vicinity of a black hole. With this motivation in mind, in this paper, we explored the properties of an interesting solution describing the magnetically charged stringy black hole. 

This paper explores the astrophysical applications of epicyclic motion to the study of QPOs and investigates the impact of electromagnetic fields on the dynamic behavior of electrically and magnetically charged particles in the accretion disk surrounding a central magnetically charged stringy black hole. By doing so, the aim is to bolster confidence in the conclusions drawn from observations pertaining to accretion disks. We studied the captured and bound orbits to understand the impact of black holes'  magnetic charges and particle charges. We observed that orbits become bound, initially captured ones, due to the effect of black hole magnetic charge $Q_m$ for the fixed magnetic charge $q_m$.  On other hand, incorporating an electric charge $q$ results in the occurrence of bound orbits either above or below the equatorial plane, depending on whether the charge is negative or positive, respectively. 

Furthermore, our analysis of particle trajectories led us to conclude that the interplay between the magnetically charged $Q_m$, electric $q$, and magnetic $q_m$ charges is a crucial factor in elucidating the behavior of bound and captured orbits in the vicinity of a black hole. Additionally, we investigated key ISCO (innermost stable circular orbit) parameters of test particles, such as $r_{ISCO}$, $v_{ISCO}$, and $\Omega_{ISCO}$. These parameters are crucial for astrophysical applications, as they play a significant role in determining the inner edge of the accretion disk. 

Our findings explicitly demonstrate that the ISCO radius increases as $Q_m$ (black hole magnetic charge) increases, suggesting that $Q_m$ functions as an attractive gravitational charge. The impact of the charge coupling parameter $g_m$ is analogous to that of $Q_m$, with a positive value resulting in a similar effect, while a negative value produces an opposing effect that leads to a reduction in the ISCO radius.  

Furthermore, our research revealed that the Keplerian frequency $\Omega_{k}$ decreases as $Q_m$ increases at larger values of $r$. However, the opposite trend occurs at smaller values of $r$, particularly in close proximity to the black hole horizon. Similarly, the orbital velocity $v$ experiences a decline due to the influence of $Q_m$. Nevertheless, both $\Omega_k$ and $v$ exhibit predominantly increasing behavior as test particles approach smaller values of $r$.

We studied the epicyclic frequencies measured by the distant observer located at spatial infinity and demonstrated that epicyclic frequencies for the distant observer decrease as a consequence of increasing black hole magnetic charge $Q_m$. By an explicit calculation, we demonstrated that the coupling parameter $g_m$ strongly affects the radial frequency $\omega_r$, i.e., it respectively increases and decreases with increasing the coupling parameter $\pm g_m$. It turns out that the coupling parameter $\sigma_m$ can only alter the latitudinal frequency $\omega_{\theta}$. 

Further, we considered the HF QPOs observed in three microquasars for black hole parameters to better fit with the model considered here. For the best fit of observed data, we demonstrated that the combined effect of the charge coupling parameters $g_m$ and $\sigma_m$ play an increasingly important role with respect to the neutral particle case. Also, to test our model, we have used the Monte Carlo simulations for the observed frequencies for three selected microquasars from where we extract  constraints on the parameters of the magnetically charged stringy black holes as well as coupling parameters describing the interaction between the particle and the black hole. To this end, we have presented the best fit values for each case separately. For the black hole mass, we find a greater value compared to the magnetic charge $Q_m$. The coupling parameters can have both positive and negative values. From the results, we clearly observe that besides the mass, the magnetic charge plays a very important role in the QPOs. 

We also explored the radiation of the accretion disk, i.e., the radial profile of the flux of the electromagnetic radiation for various combinations of $Q_m$ and $g_m$. Interestingly, we observed that the flux of the radiation respectively decreases and increases as a consequence of an increase in the value of black hole magnetic charge $Q_m$ and the charge coupling parameter $g_m$. Additionally, our findings indicate that the temperature and energy of the accretion disk are lowered as a consequence of the influence of $Q_m$ (as seen also in color map in Fig.~\ref{temperature2}). Finally, we also studied the behaviour of differential luminosity referred to as one of the important quantities delineating nature of the accretion disk, introducing it by the spectral luminosity expression as a function of $\nu$.  We found that the differential luminosity for the accretion disk also decreases slightly as a consequence of the black hole magnetic charge parameter $Q_m$, similar to what is observed for the accretion disk temperature and energy.

It is worth mentioning that the research conducted in this paper on the epicyclic motions around the magnetically charged stringy black hole, as well as the constraints on the parameters of the black hole and charge coupling parameters using the observed frequencies of three selected microquasars, is both novel and of significant astrophysical importance. This is particularly noteworthy because the model developed here to establish these constraints has not been previously introduced in the literature.

\label{Sec:Conclusion}

\section*{Acknowledgments}
S.S. and B.A. wish to acknowledge the support from Research F-FA-2021-432 of the Uzbekistan Agency for Innovative Development. 
\bibliographystyle{apsrev4-1}  
\bibliography{gravreferences,ref}

\end{document}